\documentclass{MITcsail}

\title{Knowing Unknowns in an Age of Information Overload}

\author
{Saurabh Khanna~$^{1, 2}$\footnote{Correspondence E-mail: s.khanna@uva.nl}\\
\vspace{1em} 
\normalfont{\small $^{1}$Amsterdam School of Communication Research, University of Amsterdam}\\
\normalfont{\small $^{2}$Pembroke College, University of Oxford}\\
}

\usepackage{pifont}
\usepackage{threeparttable}
\usepackage{setspace}
\newcommand{\sym}[1]{^{#1}}

\begin{document}

\maketitle
\thispagestyle{firstpagestyle} 

\begin{abstract}
The technological revolution of the Internet has digitized the social, economic, political, and cultural activities of billions of humans. While researchers have been paying due attention to concerns of misinformation and bias, these obscure a much less researched and equally insidious problem -- that of uncritically consuming `incomplete information'. The problem of incomplete information consumption stems from the very nature of explicitly ranked information on digital platforms, where our limited mental capacities leave us with little choice but to consume the tip of a pre-ranked information iceberg. This study makes two chief contributions. First, we leverage the context of Internet search to propose an innovative metric that quantifies 'information completeness'. For a given search query, this refers to the extent of the information spectrum that is observed during Internet browsing. We then validate this metric using 6.5 trillion search results extracted from daily search trends across 48 nations for one year. Second, we find causal evidence that awareness of information completeness while browsing the Internet reduces resistance to factual information, hence paving the way towards an open-minded and tolerant mindset.
\end{abstract}

\section{Introduction}
\label{introduction}

Humans are in the middle of a transition -- a transition to a life on the Internet. In the last two decades, our interactions have experienced the beginnings of a digital metamorphosis that is still unfolding \citep{lazer2009computational, lazer2020computational, hofman2021integrating}. These changes are largely driven by the technological revolution of the Internet, which has effectively digitized the social, economic, political, and cultural activities of billions of people, generating vast repositories of digital data as a byproduct \citep{lazer2020computational}. The scale of this revolution is indicated by more than 8 billion Internet searches originating every day on Google alone, which roughly corresponds to one daily search for each human living on our planet \citep{stats}. The COVID-19 pandemic arrived as a powerful catalyst for this already amplifying revolution by rapidly normalizing a `remote' lifestyle. \cite{brynjolfsson2020covid} surveyed a nationally-representative sample of the American population during the COVID-19 pandemic showing that half of individuals employed pre-pandemic were now working from remote locations. Beyond the labor market, schools and universities transitioned to remote learning too as an increase in online education led to greater Internet dependence for both students and educators \citep{daniel2020education, ali2020online}. These transitions towards a digitized lifestyle did not stay restricted to employment or learning alone, as the pandemic witnessed teenagers' daily Internet use for non-school tasks consistently exceed pre-COVID levels \citep{vogels2022teens}. These changes were also not restricted to any particular demographic, as the United States saw an overall 47\% rise in broadband Internet usage across the country \citep{brake2020lessons}.

On one hand, as the Internet transforms the way we access and share information, it has clearly enabled a democratic discourse by facilitating public participation and encouraging deliberation. The Internet has democratized access to information and made it easier for citizens to participate in democratic processes. Online platforms such as social media, blogs, and discussion forums provide opportunities for individuals to express their opinions, share news, and engage in public debates \citep{dahlberg2001internet, hague1999digital}. Moreover, the Internet has enabled new forms of digital activism and civic engagement, allowing citizens to organize, mobilize, and campaign for social and political change \citep{gerbaudo2017cyber}. It fosters deliberative democracy by providing spaces for reasoned discussion and debate. Online platforms enable individuals to engage with others who have different perspectives, leading to a more comprehensive understanding of complex issues \citep{schwartz1996netactivism, hermes2006citizenship}. Additionally, the Internet allows for real-time feedback and interaction, making it possible for discussions to evolve dynamically and respond to new information and arguments \citep{fettweis2014tactile}.

On the other hand, while this explosion in freely available online information has enabled human voices across space and time, concerns have also been raised around potential harms of the information flowing on the Internet. Scientists across disciplines have made progress studying these concerns along two themes. The first theme pertains to the propagation of \emph{misinformation}, where the information being propagated is different from the ground truth for a given context \citep{west2021misinformation, roozenbeek2020susceptibility, swire2019public}. A second theme has been the growing focus on algorithmic fairness and the propagation of \emph{bias}, wherein the information propagated not only differs from the ground truth, but also can particularly harm traditionally marginalized populations \citep{obermeyer2019dissecting, ledford2019millions, cavazos2020accuracy}. But there is a crucial loophole here. Notwithstanding the validity and the gravity of the questions addressed by these two themes, they do depend on the availability of verifiable objective truths. Given the subjectivity and diversity in opinions expressed on the Internet, the presence of verifiable objective truths is more of an exception rather than a norm \citep{vosoughi2018spread}. For instance, if a certain politician makes a claim that `People think this election was rigged!', we do not have a way to dynamically assign truth or falsehood to this statement on the spot, not unless we find at least one individual who does think the election was rigged at that moment in time. Moreover, if there is this one individual who thinks that the election was rigged, that makes the politician's statement objectively true, but does not say anything about whether the election was in fact rigged. It is extremely difficult to objectively evaluate the quality of information on the Internet when the ground truths themselves are unclear, or even nonexistent.

While we have made promising progress on this demanding task of countering misinformation and bias, we have missed out on tackling another potent and arguably equally tenuous problem -- that of being subject to severe information overloads and uncritically consuming \emph{incomplete information}. A direct consequence of our rapidly digitizing lifestyles is that our information sources are no longer restricted to our social networks in the physical world. Rather, we are inundated with information from multiple sources, with both the sources and the information they carry growing at an alarming rate. 
Estimating the exact rate at which this information is growing is difficult due to the complex and rapidly evolving nature of data generation, storage, and dissemination. 
But a seminal study by the International Data Corporation projects that the global datasphere, which encompasses all the digital data created, captured, or replicated, would grow from roughly 33 zettabytes in 2018 to 175 zettabytes by 2025.\footnote{One zettabyte is equal to $10^{21}$ bytes, or a trillion gigabytes.} This estimate represents a substantial compound annual growth rate of approximately 61\% over a five-year period \citep{rydning2018digitization}.

Given this rapidly growing volume of information on the Internet, I see two aspects governing our interactions with it. First, all information shown to us on the Internet is `ranked' by nature. In the context of web search, for instance, the \(n^{th}\) search result ranks higher than the \(n+1^{th}\) search result. In the context of social media feeds, the \(n^{th}\) post in our feed is ranked higher (and hence more visible) than the \(n+1^{th}\) one. While this ranking certainly takes into account our prior interactions with the platform, it is largely decided by recommendation algorithms acting to maximize our engagement almost entirely beyond our control \citep{guy2011social, zhou2012state}. Second, when dealing with this pre-ranked information, we as humans are severely restricted by the bounds of our own rationality. We may not want to spend time scrolling through multiple pages of search results, and clicking on what is easiest to click not only minimizes cognitive load but also saves time. In other words, we lack the mental capacities to keep up and effectively process the exponentially growing faucet of information we face everyday. Consequently, we react to this pre-ranked digital information with an extreme predilection for the tip of the iceberg, where our clicks roughly follow a power law distribution \citep{introna2000shaping}. 

This context leads us to a natural and fundamental question -- \emph{how much of the information spectrum am I seeing as I am browsing the Internet?}. In more concrete terms, from a population of \(N\) search results output for a given search query \(q\) on the Internet, how representative is viewing just \(n\) \((< N)\) search results? This is different from assessing whether the \(n\) search results are either misinformative or biased or both, but worth assessing nonetheless. The importance of this question is even more pronounced given the implications it has for human behavior. Studies have shown the rising levels of mental distraction among almost all population demographics, a large part of which is driven by the fear of missing out on what we could not see \citep{paasonen2021dependent, harris2022happiness}. Additionally, the misinformation and bias literature itself has highlighted the existence of prejudiced information in top web search results \citep{goldman2005search, yue2010beyond}, top news search results \citep{groeling2013media}, and top social media posts \citep{kulshrestha2017quantifying}. This is rather unsurprising as search algorithms are built using data that reflects historical and societal biases \citep{goldman2005search}. If the training data contains biased information, the search algorithms may perpetuate these biases, leading to unfair representation of minority groups in search results. The overall picture then is problematic as the Internet sends us pre-ranked information, a ranking which we feed sparingly off, and a ranking which could \textit{possibly} be misinformative and biased.\footnote{The emphasis on `possibly' pertains to the ambiguity we face in accurately assessing the ground truths in most situations.} This in turn can lead to what \cite{ananny2018seeing} refer to as `harms of representation', wherein digital systems end up reinforcing the subordination of certain groups along the lines of identity. 

Taken together, our failure to know how much we do not know (or `knowing unknowns') when consuming information is a critical loophole in Internet-enabled systems facilitating human discourse at an unprecedented scale. As \cite{susskind2018future} points out, if we have no control on the flow of information in our society, we have no control on our shared sense of right and wrong. From a philosophical standpoint, an intention to know unknowns is hardly a new line of questioning, but rather a centuries old one ranging from proponents like Plato \citep{cooper1997plato}, Einstein \citep{einstein1931knowledge}, and more recently Taleb \citep{taleb2007black}. But notwithstanding the fundamental nature and increased utility of answering this question in the current information overload age, it is rather surprising that this question has evaded ample research attention. We have been trying to answer `Is what I know different from the ground truth?' through research on misinformation and algorithmic bias, but are yet to adequately answer `How much do I not even know?'. Once we start approximating an answer the latter question, we would also be better placed to understand the behavioral implications of consuming partial knowledge at both the individual and the societal levels.

Motivated by this need for knowing unknowns in an age of information overload, this study seeks to address two primary objectives. First, I aim to quantify the extent of the information spectrum that is visible to individuals as they navigate the Internet. This involves assessing the proportion of accessible information relative to the totality of available content. Second, I intend to delve into the behavioral implications of having an awareness of information completeness while browsing the Internet. This aspect of the study will explore how individuals' recognition of potential information gaps can influence their online behavior, decision-making, and overall engagement with online content. By shedding light on these two objectives, this research aims to contribute to a more comprehensive understanding of our interaction with the digital landscape and provide valuable insights into the potential benefits and challenges of navigating the complex world of information in the age of information overload.

This study makes two chief contributions. First, building on information retrieval and text embedding approaches, I propose a novel metric that measures `information completeness' dynamically when one browses the Internet. In addition to being intuitive in terms of comparing low-dimensional vector representations of text, the metric is validated by assessing aspects of its distribution in 6.5 trillion web and news search results across 48 nations. Second, I find causal evidence that awareness of information completeness while browsing the Internet increases tendencies for open-mindedness, especially on account of a reduced resistance to factual information, as well as a reduced tendency for dogmatism.

The subsequent sections are organized as follows. 
Section \ref{info-ret-and-textem} provides background on the of human quest for knowledge online, and provides an overview of the evolution of Internet search over the last three decades. 
Section \ref{data-and-methods} details the approach taken to meet the study's objectives, with section \ref{approach-metrics} detailing the data and methods leveraged for the first objective of quantifying information completeness, and section \ref{approach-exp} detailing the experimental design investigating the implications of staying aware of incomplete completeness on our behavioral choices. Section \ref{results} describes results, with sections \ref{result-metrics} and \ref{result-exp} detailing findings on validation of the information completeness metric, and its behavioral implications, respectively. Section \ref{discussion} discusses the implications of our findings, as well as limitations and directions for future research.

\section{The Evolution of Internet Search}
\label{info-ret-and-textem}

The evolution of Internet search is a captivating tale of continuous innovation and adaptation, reflecting the ever-evolving needs of individuals in the digital age. Over the past three decades, search engines have played a pivotal role in transforming how we access and navigate the rapidly expanding digital universe. As the Internet has exponentially expanded, so too has the need for effective search tools to help individuals find relevant information quickly and easily. This section outlines a concise overview of the evolution of Internet search, from its humble beginnings to the sophisticated algorithms that power today's search engines.

\subsection{Early Solutions}

In the late 1980s and early 1990s, the Internet was still in its infancy, largely used by researchers, academics, and government organizations for communication and collaboration purposes. The World Wide Web, invented by Tim Berners-Lee in 1989, aimed to make the Internet more accessible and user-friendly \citep{berners2001semantic}. However, as more web pages and resources started populating the digital realm, finding relevant information became increasingly challenging. Before the advent of search engines, individuals had to rely on manually maintained lists of websites, known as directories. These directories were organized into categories and subcategories, helping individuals navigate through the growing number of websites. One of the early directory services was the Gopher protocol, developed at the University of Minnesota in 1991 \citep{anklesaria1993internet}. Gopher provided a hierarchical structure for organizing documents and resources, but its limitations quickly became apparent as the volume of online content continued to expand.

The development of Archie in 1990 marked a significant turning point in the emergence of Internet search. Archie, created by Alan Emtage, was the first search engine that allowed individuals to find specific files on public FTP sites using keywords \citep{schwartz1992comparison}. Archie's significance lay in its ability to automate the process of information retrieval, providing a glimpse into the future potential of search engines. Following Archie, several other search engines and indexes emerged, each attempting to improve upon their predecessors. Veronica and Jughead were developed as extensions of the Gopher protocol, providing keyword-based search capabilities within the Gopher system \citep{mardikian1994use, tennant1994crossing}. Around the same time, the Wide Area Information Servers (WAIS) system, developed by Brewster Kahle in 1991, allowed individuals to search through indexed databases using natural language queries \citep{livingston2007brewster}. 

As the World Wide Web grew in popularity, the need for more advanced search tools became apparent. The early to mid-1990s saw the introduction of web-based search engines such as Aliweb (1994), WebCrawler (1994), Lycos (1994), Infoseek (1994), and AltaVista (1995) \citep{seymour2011history}. These search engines used web crawlers, also known as spiders or robots, to traverse the web, indexing pages and their contents to enable keyword-based searches. The emergence of Internet search in the early 1990s laid the foundation for the rapid evolution and innovation that would follow. These early search engines, while basic compared to their modern counterparts, represented a significant leap in accessibility and user-friendliness.

\subsection{The Rise of Google}

The search engines of the mid-1990s used keyword-based algorithms to index and rank web pages, with varying degrees of sophistication. With numerous search engines vying for dominance in an increasingly competitive landscape, Google emerged as the game-changer that would ultimately revolutionize the way we search and access information online. Founded in 1998 by Larry Page and Sergey Brin, two doctoral students at Stanford University, Google was built on the premise that a search engine's ability to deliver relevant results could be vastly improved by analyzing the relationships between web pages. This insight led to the development of PageRank, a groundbreaking algorithm that assigned a numerical value to web pages based on the number and quality of incoming links \citep{page1998pagerank}. 

The primary concept behind PageRank is that a link from one web page to another can be considered as a vote or an endorsement for the linked page. PageRank views the entire web as a vast graph, with web pages as nodes and links between pages as directed edges. It assigns a numerical value or rank to each web page, reflecting its importance or authority within the web. This rank is determined not only by the number of incoming links but also by the quality of those links. In other words, a link from a highly authoritative or important page carries more weight than a link from a less significant page. Further, to account for the fact that not all links are equally significant, the PageRank algorithm introduces a damping factor, typically set to 0.85. This factor represents the probability that an individual navigating the web will continue clicking on links rather than jumping to a random page. The damping factor ensures that pages with many high-quality links are ranked higher, while pages with few or low-quality links receive a lower rank. The PageRank value for each page is calculated iteratively, with the algorithm updating the rank of a page based on the ranks of the pages linking to it. This process continues until the ranks of all pages converge, usually after several iterations. Once the PageRank values have converged, they are normalized, so the sum of all PageRank values across the entire web equals one. This normalization process allows for a more meaningful comparison between the ranks of different web pages. PageRank's unique approach to ranking web pages proved to be a significant departure from existing search engines, which primarily relied on keyword frequency and density to determine relevance. By prioritizing quality content and authoritative sources, Google was able to provide individuals with more relevant and reliable search results. This innovative approach to search quickly gained traction and catapulted Google to the forefront of the search engine market.\footnote{The PageRank algorithm is just one of many factors that present-day Google uses to determine the relevance and ranking of web pages in its search results. Over time, Google has introduced numerous updates and enhancements to its ranking algorithms, including factors such as content quality, user behavior, and social signals \citep{ziakis2019important}. While PageRank may not carry the same level of influence as it once did, it remains an important foundation of Google's search technology and a key milestone in the evolution of Internet search.}

As Google gained popularity, it continued to refine and enhance its search algorithms, incorporating additional signals for individual behavior and information quality to improve search result relevance. In addition, Google introduced a number of new features and services to its platform, including personalized search, image search, Google Maps, Google News, and Google Scholar, further solidifying its position as the leading search engine. With the introduction of AdWords (later re-branded as Google Ads) in 2000, Google developed a highly effective, targeted advertising model based on keywords and user intent \citep{lee2011google}. This pay-per-click advertising model has since become the backbone of Google's revenue stream and a dominant force in the online advertising industry.

\subsection{Modern Solutions}

In present times, artificial intelligence and machine learning have transformed the landscape of Internet search, significantly enhancing the capabilities of search engines in delivering relevant and accurate results to individuals. AI-driven search algorithms have brought a new level of sophistication and understanding to the process of information retrieval, allowing search engines to better anticipate user intent, recognize context, and provide personalized results.

One of the key challenges tackled by modern search engines is deciphering the intent behind user queries. AI-driven search algorithms employ natural language processing (NLP) techniques to analyze the structure, meaning, and context of user queries. NLP algorithms can identify synonyms, variations, and related terms, enabling search engines to provide relevant results even when the exact keywords used in the query do not appear in the content \citep{yue2012analysis}. 
Semantic similarity using text embeddings can play a crucial role in improving the accuracy and relevance of search results in Internet search engines. By considering the meaning and relationships between words and concepts, rather than simply matching keywords, search engines can provide more contextually accurate and comprehensive results \citep{xia2019efficient, wilks2009natural}. I leverage this approach to build a metric for information completeness in section \ref{approach-metrics}.
NLP techniques can also help search engines disambiguate queries by analyzing the context in which words are used, ensuring that the most relevant interpretation is applied \citep{sangers2013semantic, chowdhary2020natural}. By understanding the nuances of human language, search engines can more accurately interpret user intent, delivering search results that closely align with the their needs.

Machine learning has further enabled search engines to learn from vast amounts of data and identify patterns and trends in individual behavior. By analyzing factors such as click-through rates, time spent on a page, and bounce rates, machine learning models can gain insights into individual preferences and the effectiveness of search results. This information can then be used to refine search algorithms and improve result relevancy over time. These advancements also play a key role in personalizing search results based on individual individual preferences, search history, location, and device type. Machine learning models can continually adapt and refine their understanding of individual behavior and preferences, enabling search engines to deliver highly personalized and targeted results \citep{yoganarasimhan2020search, bok2022personalized}.

Modern search engines, despite their impressive capabilities and advanced algorithms, may not always provide us with a complete picture of the information available on the Internet. As elaborated in section \ref{introduction}, this limitation can be attributed not only to inherent biases present in the algorithms, but also potentially beneficial attributes like highly personalized results. Personalization tailors search results based on an individual's search history, location, and preferences, which can naturally lead to the exclusion of potentially relevant information that falls outside of these parameters. Filter bubbles, a byproduct of personalization and recommendations based on collaborative filtering, occur when individuals are exposed primarily to content that aligns with their existing views and interests, inadvertently limiting their exposure to diverse perspectives and alternative information sources \citep{bruns2019filter, schafer2007collaborative}.\footnote{Collaborative filtering is a technique used in recommendation systems to provide personalized suggestions based on the preferences and behavior of similar individuals. It operates on the principle that individuals who have exhibited similar preferences in the past are likely to have similar preferences in the future.\citep{herlocker2004evaluating, schafer2007collaborative}} Furthermore, biases in the algorithms themselves, arising from the training data or the developers' perspectives, can skew search results in favor of certain types of content. Consequently, while modern search engines have significantly enhanced the search experience, individuals must remain cognizant of these limitations and actively seek diverse sources to obtain a comprehensive understanding of the vast digital landscape.

\section{Data and Methods}
\label{data-and-methods}

While the question of knowing unknowns when browsing the Internet has largely evaded researchers' attention, our digitizing lifestyles have now provided us with both data and methods to effectively approximate an answer to this question. In line with the first objective of this study, I will leverage natural language processing and information retrieval tools to develop a metric assessing completeness of information as we navigate the Internet, and validate it using 6.5 trillion raw Internet search results collected from 48 nations over a one year period. I then achieve the second objective by running a randomized experiment assessing the effects of information completeness (as defined by aforementioned metric) on 876 participants' browsing behavior, as well as on their open-mindedness towards novel but valid information on the Internet. In meeting these objectives, this study also presents a prototype of an experimental open-source web search platform -- \href{https://github.com/sonder-labs/sonder}{Sonder} -- that can dynamically report information completeness scores as one searches the Internet.

\hypertarget{metrics}{%
\subsection{Measuring Information Completeness}\label{approach-metrics}}

My approach for measuring information completeness builds on natural language text embedding techniques and information retrieval algorithms leveraged traditionally by Internet search engines. In the last five years, information retrieval in web search has started building on text embedding approaches, where a given piece of text (say, a web search query like `floods in Pakistan') can be represented as a vector in a low-dimensional embedding space. Text embeddings, also known as word embeddings or vector representations of text, are a powerful technique that enable the conversion of textual data into a numerical format that can be easily processed and analyzed by algorithms. By representing words, phrases, or even entire documents as vectors in a high-dimensional space, text embeddings capture the semantic meaning and relationships between words in a way that preserves their contextual information \citep{levy2014dependency}. This approach has facilitated the development of advanced NLP models and applications, such as sentiment analysis, document classification, and machine translation. Text embeddings are typically generated using unsupervised learning algorithms like Word2Vec \citep{goldberg2014word2vec}, GloVe \citep{pennington2014glove}, or FastText \citep{joulin2016fasttext}, which analyze large corpora of text and learn to associate words based on their co-occurrence patterns in the data. These algorithms create dense vector representations, wherein semantically similar words are positioned close together in the vector space, allowing for the calculation of similarity scores or the identification of analogies between words \citep{kusner2015word}.

Transformers are the state-of-the-art NLP tools at present for tasks such as question answering, language modeling, and summarizing articles. Introduced by \cite{DBLP:journals/corr/VaswaniSPUJGKP17} in 2017, the transformer model is built upon the concept of self-attention, a mechanism that allows the model to weigh the importance of different words in a sentence or sequence when processing textual data. This approach enables transformers to effectively capture long-range dependencies and context information, overcoming the limitations of previous sequential models. Recurrent neural networks (RNNs) such as long short-term memory networks (LSTMs) often struggle with issues related to vanishing gradients and computational inefficiency \citep{irie2019language, sherstinsky2020fundamentals}. Further, transformers work within a set context size, while LSTMs have memory cells designed to capture long-term information. However, in real-world scenarios, LSTMs often find it challenging to do this effectively \citep{yu2019review}. A potential disadvantage of using transformers, though, is that they require large amounts of data \citep{hassani2021escaping}. But that was not a concern in this study given the extensive corpus of 6.5 trillion Internet searches collected over a year.
Advanced language models like transformers have thus introduced contextualized embeddings, which take into account not only the co-occurrence patterns of words but also the specific context in which they appear, resulting in more accurate and nuanced representations of text.
Additionally, transformers can employ position encoding to inject information about the relative positions of words within a sequence, preserving the inherent order of language data.
Transformers like BERT \citep{DBLP:journals/corr/abs-1810-04805}, RoBERTa \citep{DBLP:journals/corr/abs-1907-11692}, and the original transformer itself \citep{DBLP:journals/corr/VaswaniSPUJGKP17} have been leveraged to complete a variety of tasks by computing word-level embeddings.

That said, a task like semantic comparisons across web search results requires a strong sentence-level understanding, and using ordinary word-level transformers can often become computationally infeasible. In order to handle sentence level embeddings, we can use a modification of the standard pre-trained BERT network that uses Siamese and triplet networks to create sentence-level embeddings for several sentences, that can then be compared using a traditional similarity metric like cosine-similarity, making semantic search across a large number of sentences feasible \citep{DBLP:journals/corr/abs-1908-10084}.
Let us consider the context of searching for information on a traditional web search engine. For a given web search query \(q\) generating a total of \(N\) search results, `relevance' of a single search result \(r_i (i \in N)\) can be computed as the semantic similarity between the vector representation of the query text (say, $\vec{q}$) and the vector representation of the search result text (say, $\vec{r_i}$) in the embedding space, an approach first demonstrated by Microsoft's Deep Semantic Similarity Model \citep{palangi2016deep}. As shown by \cite{palangi2016deep}, the semantic similarity itself can be calculated using established methods like cosine similarity between vectors $\vec{q}$ and $\vec{r_i}$. The results can then be sorted in descending order of these cosine similarities to form an explicit ranking of relevant search results.

\begin{equation}
\label{eq1}
 \cos(\vec{q},\vec{r_i})= \frac{\vec{q} . \vec{r_i}}{\|\vec{q}\| \|\vec{r_i}\|}
\end{equation}

For a given query $q$, let us also consider the complete set of results returned across pages till pagination ends, as the corpus $C$ of results for that query. A shortcoming of the approach highlighted above is that a sole focus on `relevance' might get us close to the information we seek (i.e. our query $\vec{q}$ itself), but it ignores the entire breadth of information that exists out there in the complete corpus of web search results (say $\vec{C}$).

\begin{center}
Query ($\vec{q}$) 
$\xrightarrow[\text{\checkmark}]{\text{What I want}}$ 
Search result ($\vec{r_i}$)  
$\xleftarrow[\text{\ding{55}}]{\text{What exists}}$ 
Corpus ($\vec{C}$)
\end{center}

\begin{figure}[H]
\centering
\includegraphics[scale=0.25]{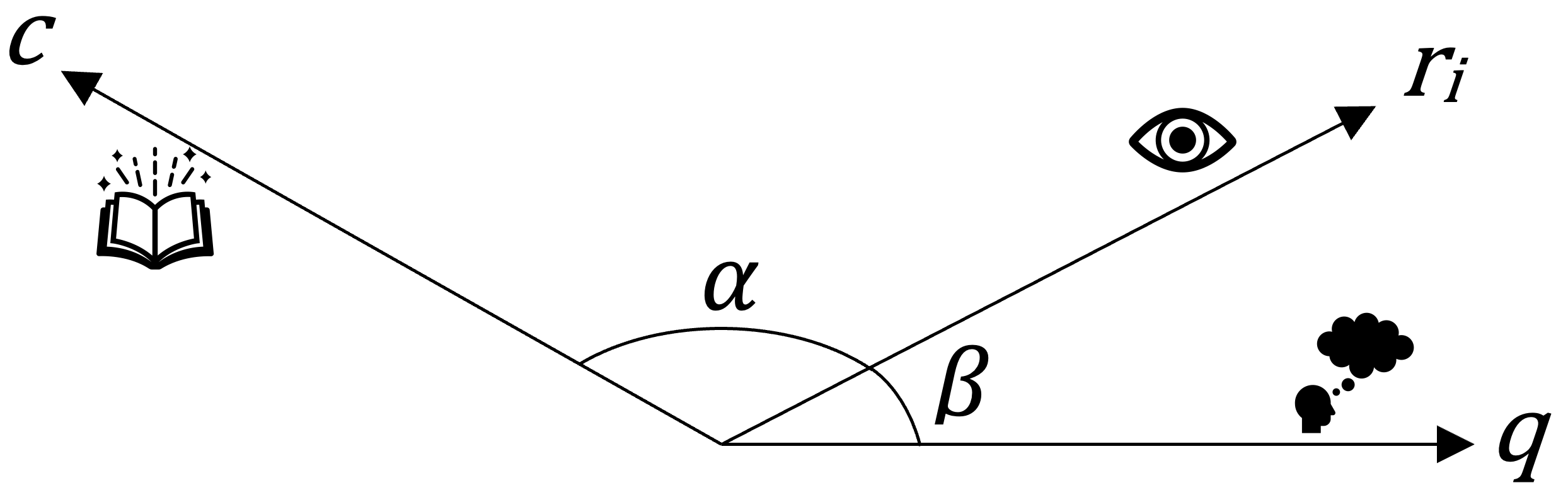}
\caption{A simplified low-dimensional vector representation of information relevance ($= cos(\beta)$) and information completeness ($= cos(\alpha)$). $\vec{q}$, $\vec{r_i}$, and $\vec{c}$ are the search query vector, search result vector, and the corpus vector respectively.}
\label{fig1}
\end{figure}

My approach builds on this shortcoming. For a given query $\vec{q}$, I alternatively consider the semantic similarity between each search result viewed $\vec{r_i}$, and the complete corpus of results $\vec{C}$. This metric essentially tells us how semantically similar a given search result $\vec{r_i}$ is to the whole corpus of search results $\vec{C}$, hence acting as a measure of information completeness -- $I_{completeness, i}$ -- for the result $r_i$. See figure \ref{fig1} below for details, where the cosine of $\beta$ refers to relevance as defined in previous information retrieval literature comparing the query vector $\vec{q}$ and the result vector $\vec{r_i}$. In my approach, the cosine of $\alpha$ refers to the information completeness metric, where each result vector $\vec{r_i}$ is compared with the corpus vector $\vec{C}$.

\bigskip

\bigskip

The embedding vector for the corpus $\vec{C}$ itself can be generated using an (ideally weighted with weights $w_i$) aggregate of all $\vec{r_i}$ vectors. Weighting can help discount results coming from websites that are either entirely unrelated to the search query \(q\) or consistently misinformative click bait. One such set of weights can be obtained through domain-level page ranks donating trustworthiness of the search result domain on a continuous scale \citep{page1998pagerank}.

\begin{equation}
  I_{completeness, i} = \cos(\vec{C},\vec{r_i}) = \frac{\vec{C} . \vec{r_i}}{\|\vec{C}\| \|\vec{r_i}\|}\text{; where } \vec{C} = \sum_{i=1}^{N} w_i \vec{r_i}
\end{equation}

There are two benefits of using $I_{completeness, i}$ as a metric for information completeness. The first benefit arises from the cumulative nature of information consumption on the Internet. When viewing the $N$ ranked results for a given search query $q$, one views the first result, then the second result, then the third, and so on till $n\text{ }(\le N)$ search results. In line with this pattern of information consumption, it is possible to compute a single cumulative $I_{completeness, n}$ score for all of $n\text{ }(\le N)$ search results viewed (instead of reporting a distinct score for each visible search result). This is useful in practical terms too, as one would be more interested in the overall information completeness achieved when one has viewed \textit{till} a particular search result (for instance, the first page of search results), as opposed to the information completeness of each individual search result on the first page. This cumulative formulation of information completeness till $n$ results have been viewed can be seen in Equation \ref{eq4}. 

\bigskip

\begin{equation}
\label{eq4}
  I_{completeness, n} = \cos(\vec{C}, \sum_{i=1}^{n} \vec{r_i}) = \frac{\vec{C} . \sum_{i=1}^{n} \vec{r_i}}{\|\vec{C}\| \|\sum_{i=1}^{n} \vec{r_i}\|}\text{; where } \vec{C} = \sum_{i=1}^{N} w_i \vec{r_i}
\end{equation}

\bigskip

\begin{figure}[H]
\centering
\includegraphics[scale=0.75, height=0.4\textheight]{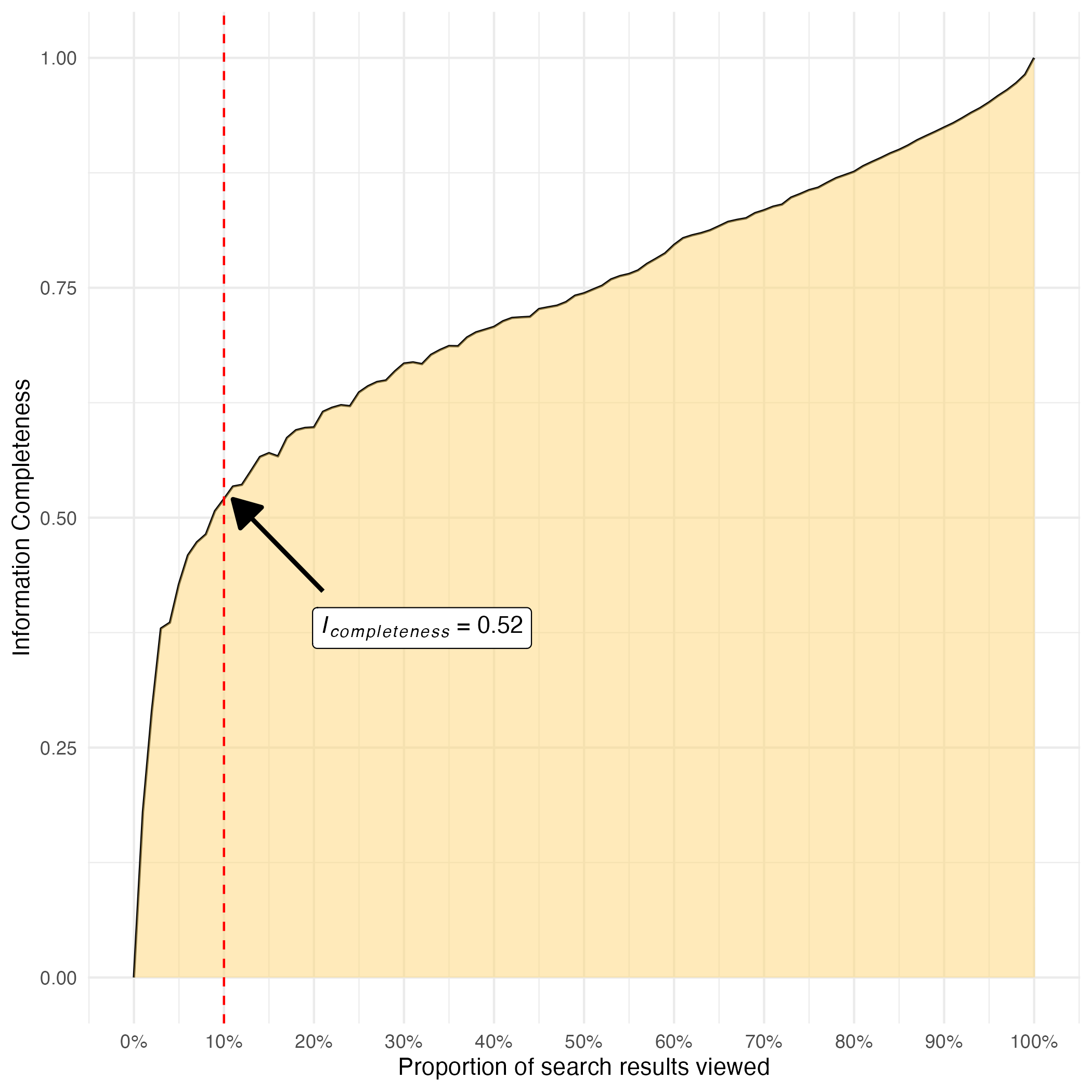}
\caption{Information completeness curve with completeness rising as more results are viewed. The dashed red line shows that the level of information completeness achieved when 10\% of the search result corpus has been viewed.}
\label{fig2}
\end{figure}

Moreover, for a given query $q$, we can plot this cumulative $I_{completeness, n}$ against the proportion of search results viewed, hence generating an `information completeness curve' (See Figure \ref{fig2}). The value of $I_{completeness, n}$ starts at 0 when nothing has been viewed (i.e. $n = 0$), and ends at 1 when all available search results have been viewed (i.e. $n = N$). The area under the curve varies directly with how efficient the entire search is in terms of information completeness. One can note that the area under the curve is also mathematically equivalent to the mean of individual $I_{completeness, n}$ values as $n$ rises from 0 to $N$.

As a second benefit, the $I_{completeness}$ metric provides us with a mathematically sound alternative for maximizing both information relevance and information completeness, hence avoiding trade-offs between two important objectives when browsing the Internet. In fact, when the search query vector \(\vec{q}\) and entire result corpus vector \(\vec{C}\) are closely aligned with each other, each search result vector \(\vec{r_i}\) is spared making a trade-off between two obtuse vectors. This is intuitive as a strong alignment between the search query \(\vec{q}\) and the whole corpus of search results \(\vec{C}\) is indicative of a good population of search results for that query, consequently leaving reduced chances of information lost through a divergent subset of search results. On the other hand, if the query and corpus vectors are not aligned, the $I_{completeness}$ metric makes it is possible to choose whether relevance or completeness matters more to the viewer, and the results can be re-ordered by what the viewer values more. To operationalize this choice for the viewer, we can assign a score \(S_i\) to each search result $r_i$ based on a balance between search result relevance (determined by $I_{relevance, i}$) and search result completeness (determined by $I_{completeness, i}$). Here, \(\lambda\) helps the participant take control of the balance between relevance and completeness.

\begin{equation}
    S_i 
    = \lambda {I_{i,completeness}} + (1 - \lambda) I_{i,relevance}
\end{equation}

where \(\lambda_i \in [0, 1]\). A higher $\lambda$ values information completeness more, whereas a lower $\lambda$ values information relevance more. The search results can finally be viewed in descending order of overall $S_i$ scores. 

\bigskip
\noindent
\textit{Data}
\medskip

\begin{figure}[H]
\centering
\includegraphics[width=\textwidth,height=0.5\textheight]{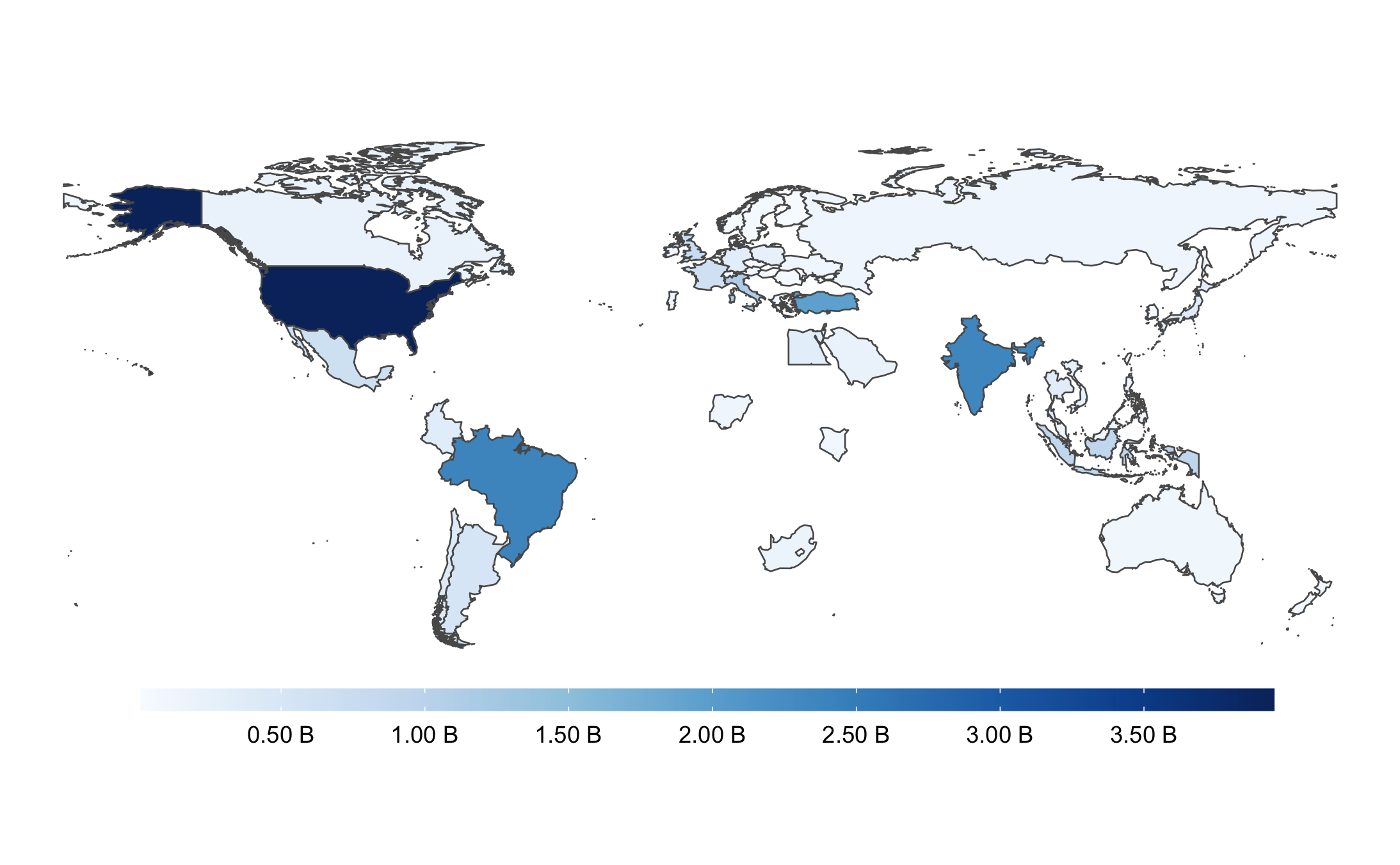}
\caption{Total search volume fetched every day across 48 nations. Color intensity in map varies directly with the raw search result volume (units in billions).}
\label{fig3}
\end{figure}

In an effort to assess the efficacy of my information completeness metric, I constructed and leveraged an extensive dataset consisting of 6.5 trillion raw Internet search results. These results were obtained by examining daily trending Google search queries across a diverse set of 48 nations, spanning six continents, for a duration of one year\footnote{The Internet search data was collected from November 16, 2021 to November 15, 2022.}. Throughout this period, I systematically extracted trending search queries for each calendar day using daily search queries released by Google Trends.\footnote{\href{https://trends.google.com/trends/trendingsearches/daily?geo=US&hl=en-US}{Google Trends} platform provides insights into search query patterns and volumes over time, and has become increasingly valuable for researchers to obtain a comprehensive understanding of the intricate relationships between online behavior and real-world events. By analyzing trends in Internet search data, researchers can gauge public interest, awareness, and sentiment regarding specific topics, events, or phenomena \citep{choi2012predicting}. This information has been utilized in a wide range of research areas, such as tracking the spread of infectious diseases \citep{nuti2014use}, predicting economic indicators \citep{woloszko2020tracking}, studying the impact of political events \citep{mellon2014internet}, and examining consumer behavior \citep{carriere2013nowcasting}.} 
Subsequently, for each search query, I amassed raw web search and news search results by initiating incognito searches from IP addresses located in each of the 48 nations under study. On a daily basis, I successfully gathered data pertaining to 57.6 million searches conducted across the 48 nations. Further, each search query produced a median of 320 search results until the Google pagination reached its limit (i.e., the point at which it was no longer possible to navigate to subsequent pages to obtain additional results). In aggregate, the entire process effectively generated 18 billion data points (i.e. search result texts) every day. Figure \ref{fig3} offers a geographic representation of the Internet search data extracted from the 48 nations. The regional color intensities in this figure serve to effectively illustrate the volume of search results acquired for each nation.

With this extensive dataset at my disposal, I then proceeded to evaluate the information completeness metric $I_{completeness}$ for each search query under investigation. This evaluation was conducted for each nation on a daily basis and cumulatively over the entire one-year period. In the results section \ref{result-metrics}, I aggregate these information complete scores to the country level, and examine the variation of these scores with the extent of media freedom in the country assessed using data available from Reporters Without Borders \citep{reporters, leeson2008media, becker2007evaluation}. These nationwide comparisons are interesting because from a political standpoint, governments might control information to maintain power, suppress opposition, or manipulate public opinion. This could involve censoring content that criticizes the government, challenges the status quo, or advocates for political change. For instance, they may block websites or social media platforms used to organize protests, or filter out news that paints the government in a negative light. The goal would be  to shape the narrative that citizens are exposed to, thereby influencing their perceptions and attitudes \citep{de2020internet}. Moreover, nations might control online information to preserve national identity, cultural values, or moral standards. This could involve blocking content that is viewed as being in conflict with traditional values or norms, or that is seen as a form of cultural imperialism \citep{bowrey2005law}. From a social perspective, controlling internet information might be seen as a way to protect societal harmony. This can involve the censorship of content that is considered harmful, offensive, or divisive. For example, a nation might restrict access to information that promotes hate speech, violent extremism, or misinformation in an attempt to maintain social cohesion and public safety \citep{wok2017internet}. Furthermore, in the era of cyber warfare, nations may also control internet information to protect national security. This might involve monitoring internet traffic to guard against cyber threats, or blocking access to sensitive information that could be exploited by foreign adversaries \citep{lu2018internet, abomhara2015cyber}. Therefore, the issue of country-level Internet freedom raises important questions about the balance between state control and individual freedoms in the digital age, and can be a way to examine the efficacy of our information completeness metric.

\subsection{Implications of Incomplete Information Awareness}
\label{approach-exp}

My first objective was aimed at defining and validating a metric to assess information completeness on the Internet. Building on that, my second objective is aimed at understanding the implications of awareness about information completeness on human behavior. I run a randomized experiment assessing the affects of viewing information completeness scores on participants' browsing behavior and open-mindedness towards novel and conflicting view points. 
This section details the experimental design of the study -- the sampling, the randomization protocol, the pre-treatment and the post-treatment surveys, the outcomes, and the analysis specifications.

I recruited a sample of 876 adult US participants through Prolific, a platform that allows researchers to publish Human Intelligence Tasks and compensate subjects who choose to participate \citep{palan2018prolific}. Prolific has a diverse pool of participants from various demographic backgrounds, which allows researchers to select the exact demographic they're interested in. This is useful as one could limit the availability of the task to respondents who meet certain filer, which in my case were -- i) English-speaking participants, ii) aged 18 and above, and iii) living in the United States. Each participant is then assigned a username (which is the same as their unique Prolific ID) and a password to log into \textit{Sonder} -- the internet search platform developed and tailored for this experiment. The experiment duration lasts 40 minutes, including a 5-minute pre-test, 30-minute interaction with the search platform, and a 5-minute post-test.

Once a participant has logged in, the search platform randomly assigns them to one of the two experiment arms -- one treatment arm (browsing search results with visible information completeness scores) and one control arm (browsing search results as usual without any information completeness scores). The participants in both arms start with a pre-test questionnaire, which collects information on -- i) the participants' demographic variables (their age, gender, ethnicity, education, annual income, domicile, and religiosity), and ii) their responses on a AOT7 survey assessing open-minded thinking. Derived from \cite{haran2013role}, this 7-item survey is a psychological tool used to measure the extent to which individuals are willing to consider and engage with ideas that are different from their own pre-existing beliefs and opinions. This scale is often used in research to assess a person's cognitive style, especially in relation to their willingness to update their beliefs in the face of new evidence or arguments. All items are scored on a 6 point scale from $-$3 (strong disagreement) to 3 (strong agreement). Scores can then be aggregated across the seven items to create a single AOT7 score. An extended 17-item version of this scale is considered as an outcome in the posttest phase, and will be described ahead with the outcomes being assessed.\footnote{I further collect information on participants opinions on 5 items from the social and economic conservatism scale (SECS) scale proposed by \cite{everett201312}.} 

The assignments to each experimental arm are equally likely in probability and we have 434 participants in the treatment arm, and 442 participants in the control arm. The treatment assignment satisfies balance tests on pre-treatment participant characteristics as seen in Table \ref{tab1}. We see a statistically significant difference in the proportion of urban participants across the arms, but the magnitude of this difference (0.03) is small.

\bigskip

\begin{table}[H]
\centering
\begin{scriptsize}
\begin{threeparttable}
\begin{tabular}{@{}lccc@{}}
  \toprule
                              & Treatment Group & Control Group & Difference (SE)         \\ \midrule

  White                       & 0.69            & 0.66          & 0.03 (0.02)         \\
  Female                      & 0.45            & 0.47          & -0.02 (0.02) \\
  Age                         & 28.74           & 27.80         & 0.94 (1.04)        \\
  College graduate            & 0.43            & 0.47          & -0.04$\sym{*}$ (0.02)         \\
  Annual Household Income $\ge$ \$60k & 0.58    & 0.56          & 0.02 (0.02)         \\
  Urban Domicile              & 0.82            & 0.85          & -0.03$\sym{**}$ (0.01)        \\
  Democrat                    & 0.57            & 0.55          & 0.02$\sym{*}$ (0.01) \\
  AOT7 Score (Standardized)     & 0.02            & -0.02          & 0.04 (0.07)         \\\\
  \textit{N}                  & 434             & 442           &                     \\
  \bottomrule
\end{tabular}
\begin{tablenotes}
    \item Treatment and control columns report group means. The last column reports difference in means with standard errors in parentheses. AOT7 scores are standardized across the participating populations. Significance levels: $\sym{***}$ $p<0.01$, $\sym{**}$ $p<0.05$, $\sym{*}$ $p<0.10$.
\end{tablenotes}
\end{threeparttable}
\caption{Balance table of pre-treatment covariates across the treatment and control groups.}
\label{tab1}
\end{scriptsize}
\end{table}

\begin{figure}[H]
\centering
\includegraphics[width=\textwidth]{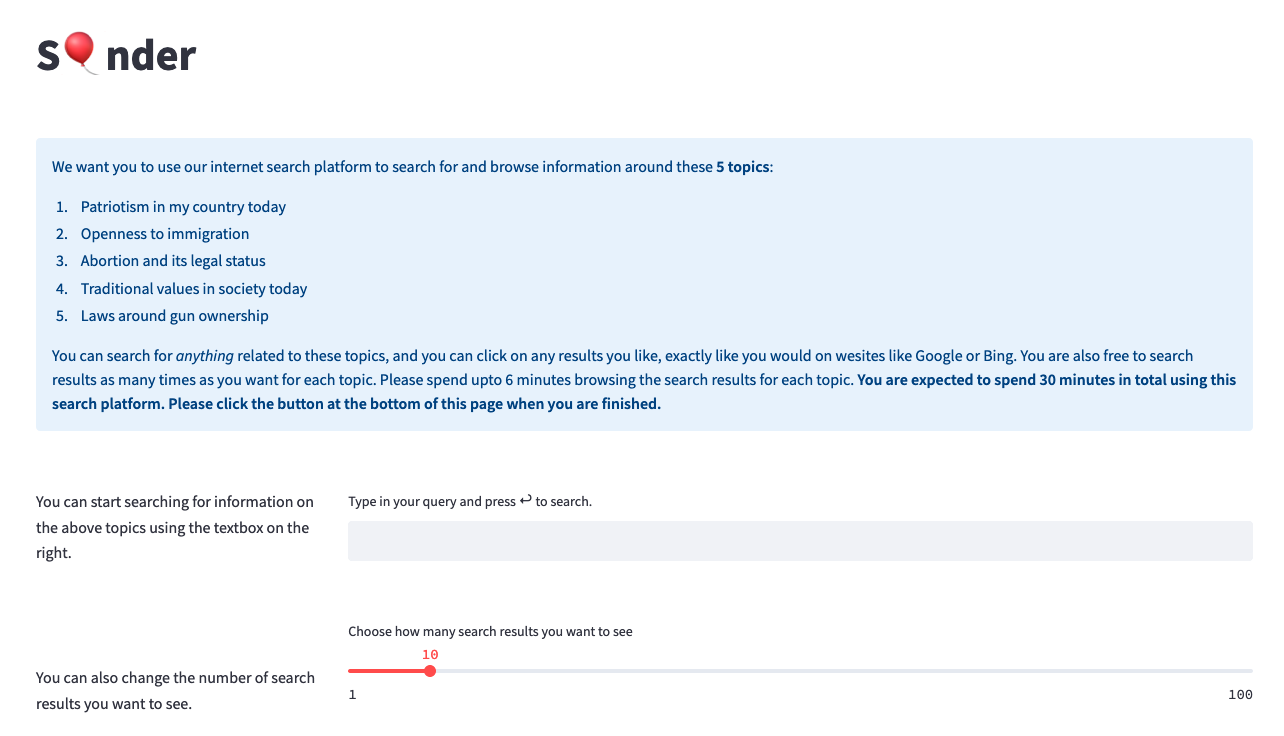}
\caption{Search platform view before any query is entered. Instructions for study participants are noted at the top.}
\label{figdemo1}
\end{figure}

\begin{figure}[H]
\centering
\includegraphics[width=\textwidth,height=0.4\textheight]{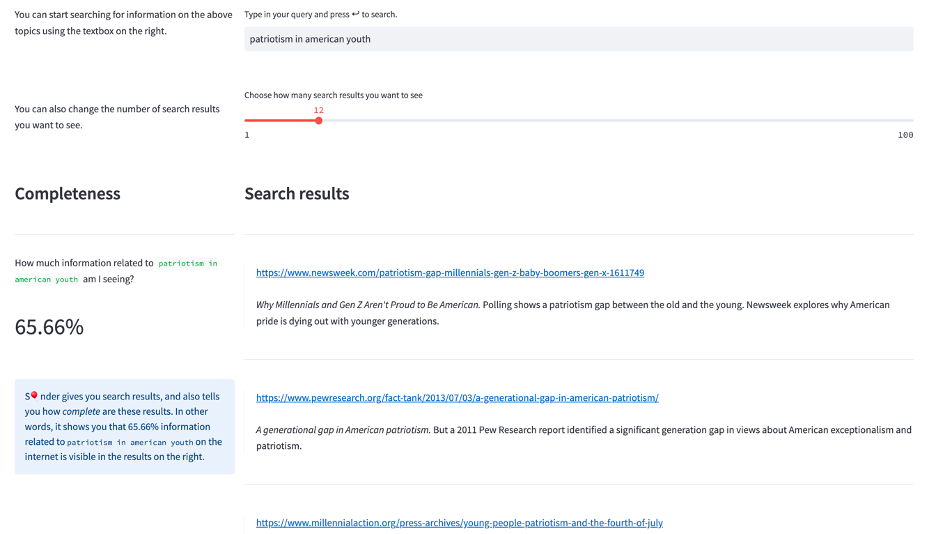}
\caption{Search platform view for \textit{treatment} group participants when they enter a search query (e.g., `patriotism in American youth'). An information completeness score is provided.}
\label{figdemo2}
\end{figure}

\begin{figure}[H]
\centering
\includegraphics[width=\textwidth,height=0.43\textheight]{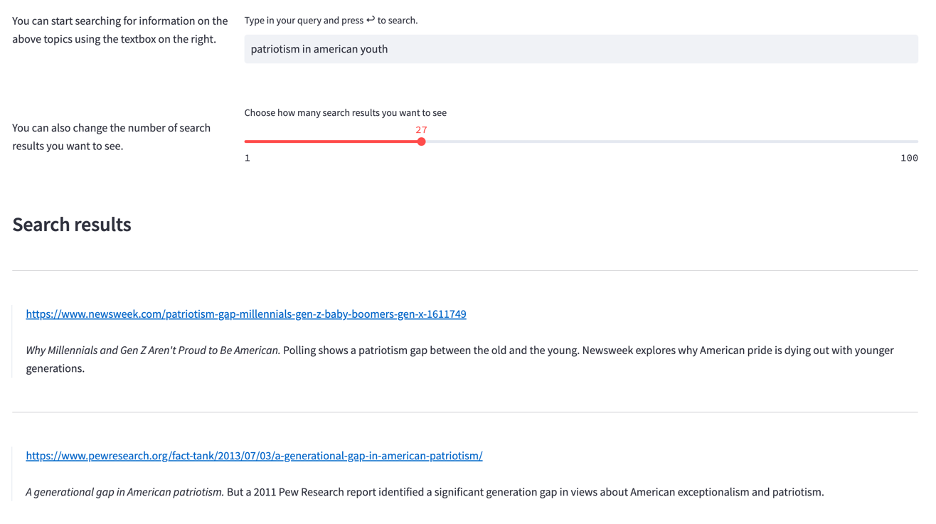}
\caption{Search platform view for \textit{control} group participants when they enter a search query. No information completeness score is provided along with the search results.}
\label{figdemo3}
\end{figure}

After completing the pre-test, participants move to interact with the search platform in their respective treatment arms.
Search results in both experimental arms are ranked exactly in line with the ranking visible in incognito searches on Google search for the same queries. 
Each participant across the treatment and control arms is provided with 5 topics that they can use for web search on the platform. The five topics for each participant pertain to broad issues relevant for assessing open-mindedness and responses to potentially conflicting information: `Patriotism in my country today', `Openness to immigration', `Abortion and its legal status', `Traditional values in society today', and `Laws around gun ownership'. 
Participants can type and search for any query related to these topics, exactly like they would on search engines like Google or Bing. For each topic, they are also free to type and search as many times as they wish. The default setting is set to show 10 search results, but the participants can use a slider to see upto 100 search results.
A visual layout of the instructions and search controls available on the platform can be seen in  Figure \ref{figdemo1}. 
The visual setup with search results (after a search query has been made) for the treatment group and the control group can be seen in Figures \ref{figdemo2} and \ref{figdemo3} respectively, where the first figure includes information completeness scores that are only visible to the treatment group.
After a 30-minute interaction with the search platform, a post-test is administered to both groups which observes the effects of the intervention on outcomes described ahead.

The experiment considers two outcomes assessing different aspects of participant behavior. The first outcome $O_1$ examines a participant's openness to novel or opposing view points when searching for information on the Internet. To evaluate this outcome, I leverage an extended version the Actively Open-minded Thinking (AOT) instrument used extensively in personality psychology literature \citep{stanovich2007natural}. As with the AOT7 scale used in the pretest, the 17-item AOT17 scale is assesses a person's cognitive style with regards to their willingness to update their beliefs in the face of new evidence or arguments \citep{svedholm2018actively}.
According to \cite{baron2000thinking}, `active' in AOT refers to not waiting for these things to happen but seeking them out, `open-minded' refers to the consideration of new possibilities, new goals, and evidence against possibilities that already seem strong, and good `thinking' refers to search that is thorough in proportion to the importance of the question, confidence that is appropriate to the amount and quality of thinking done, and fairness to other possibilities than the one we initially favor. 
\cite{stanovich2007natural} argue that individuals who score high on AOT scales are generally less prone to confirmation bias, a cognitive bias where people favor information that confirms their existing beliefs and disregard or devalue information that contradicts them.
As a thinking disposition, AOT assesses traits well aligned with our outcome $O_1$. Further, \citep{svedholm2018actively} highlight that the AOT17 is a multidimensional
trait with four distinct dimensions -- two of
them concerned with knowledge (a lack of dogmatism and an openness to facts even if they contradict one’s previous views) and two concerned with individuals (a liberal attitude towards people and a refusal to judge others for their opinions). Building on this work,
I use the AOT17 to assess openness among four dimensions -- fact resistance, dogmatism, liberalism, and belief personification. All survey items are scored on a 6 point scale from $-3$ (strong disagreement) to 3 (strong agreement), and can be seen in Table \ref{tab2}. The item scores can be aggregated within dimensions to create a dimension-level score, as well as overall to create a single AOT17 score ($\in [-3, 3]$). I further standardize this aggregated score across all participants while reporting experiment results in section \ref{result-exp}.

\bigskip

\begin{table}[H]
\centering
\resizebox{\textwidth}{!}{%
\begin{tabular}{@{}ll@{}}
\toprule
Dimension                               & Item                                                                                                             \\ \midrule
\multirow{5}{*}{Fact resistance} & \begin{tabular}[c]{@{}l@{}}
One should disregard evidence that conflicts with your established beliefs. (R)\end{tabular} \\
                                        & It is important to persevere in your beliefs even when
evidence is brought to bear against them. (R)             \\
                                        & Certain beliefs are just too important to abandon no matter
how good a case can be made against them. (R)        \\
                                        & Beliefs should always be revised in response to new information or evidence.                                     \\
                                        & People should always take into consideration evidence that goes against their beliefs.                           \\
                                        &                                                                                                                  \\
\multirow{6}{*}{Dogmatism}              & I believe that loyalty to one’s ideals and principles is more
important than “open-mindedness”. (R)              \\
                                        & I believe that the `new morality' of permissiveness is no
morality at all. (R)                                   \\
                                        & Of all the different philosophies which exist in the world there
is probably only one which is correct. (R)      \\
                                        & I think there are many wrong ways, but only one right way, to
almost anything. (R)                               \\
                                        & I believe letting youth hear controversial speakers can only confuse and mislead them. (R)                    \\
                                        & I believe we should look to our religious authorities for
decisions on all moral issues. (R)                         \\
                                        &                                                                                                                  \\
\multirow{3}{*}{Liberalism}             & I consider myself broad-minded and tolerant of other people’s lifestyles.                                        \\
                                        & A person should always consider new possibilities.                                                               \\
                                        & I believe that the different ideas of right and wrong that people in other societies have may be valid for them. \\
                                        &                                                                                                                  \\
\multirow{3}{*}{Belief personification} & There are a number of people I have come to dislike because of the things they stand for. (R)                        \\
                                        & I tend to classify people as either for me or against me. (R)                                                       \\
                                        & I feel anger whenever a person stubbornly refuses to admit they are wrong. (R)                               \\ \bottomrule
\end{tabular}%
}
\caption{AOT17 survey items assessing open-mindedness. Items flagged with (R) are reverse coded. All items are scored on a 6 point scale from $-3$ (strong disagreement) to 3 (strong agreement), and can be aggregated within dimensions to create a dimension-level score, as well as overall to create a single score.}
\label{tab2}
\end{table}

\bigskip

The second outcome $O_2$ explicitly examines participants' click behavior. In particular, this outcome sheds light on how much an individual is willing to go beyond the tip of the iceberg when searching for information on the Internet. In concrete terms, for each search query made by a participant, the search platform will show a total of $n$ $(\le 100)$ search results, where each search result can be represented as $r_i$ $(i \in [1, 100])$ as shown previously. We measure this outcome in three ways -- i) log the furthest ranked search result $r_i$ clicked on for each query, and extract index $i$ as an indicator of how far the participant went down the list of search results, ii) log the number of search results clicked on for each query as $n_q$, and iii) log the completeness $I_{com,ij}$ of each search result clicked. We then aggregate these indices across all five topics searched for by the participant to generate three participant level measures for this outcome.

In line with the two outcomes described above, my experimental analyses are informed by two main hypotheses. My first hypothesis is that when participants seek information on the Internet, knowing how complete their information is makes them more open to new and conflicting view points (outcome $O_1$). The second hypothesis relates to the browsing behavior of participants, and considers that the presence of information completeness makes them view more number of, more lower-ranked, and more complete search results (outcome $O_2$). The analysis specification for a given participant $j$ can be seen below:

\begin{equation}
 O_{j} = \beta_0 + \beta_1 I_{com,j} + \beta_2 X_j + \epsilon_j
\end{equation}

where Treatment \(I_{com,j}\) is viewing the information completeness metric shown along with your search results, $O_{j}$ is outcome $O_{1,j}$ or $O_{2,j}$, and $X_j$ refers to pre-treatment participant characteristics used as regression controls.

\section{Results}
\label{results}

\subsection{Measuring Information Completeness}\label{result-metrics}

\begin{figure}[H]
\centering
\includegraphics[scale=0.85]{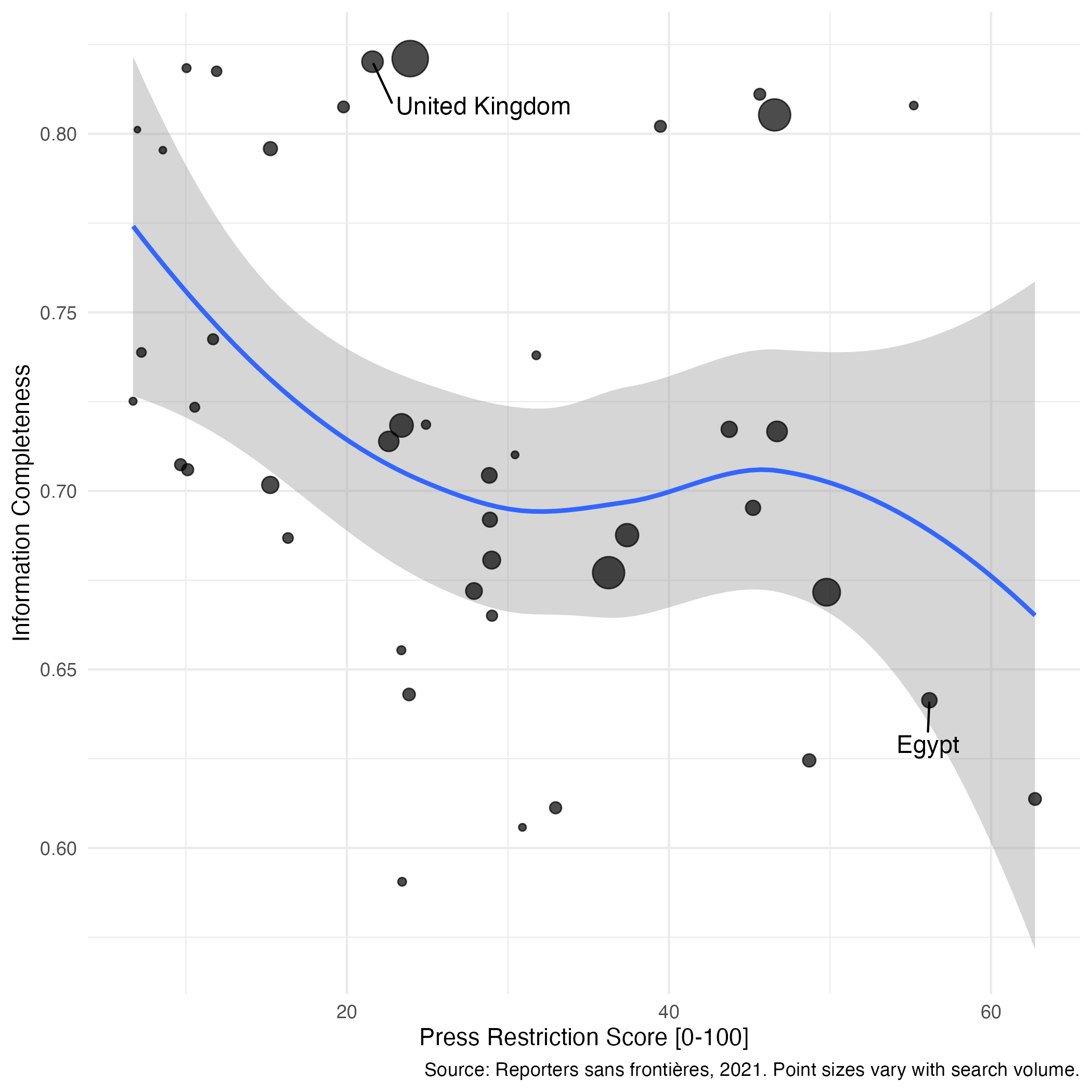}
\caption{Variation in Information Completeness with Press Restriction Scores. The size of each depicted point refers to the search volume driven by each country.}
\label{fig4}
\end{figure}

I start with results from validation checks for my approach measuring information completeness on the Internet for the 6.5 trillion web search results collected across 48 countries for one year. For every search query under consideration in our data, I generate an information completeness curve, similar to the one seen previously in Figure \ref{fig2}. As noted earlier in section \ref{approach-metrics}, the area under an information completeness curve denotes how efficient those searches are in terms of showing us more complete information in the beginning (i.e. how representative the top results are of the entire corpus). It is interesting to see the information completeness curves aggregated to the country level, because national governments' regulation of digital media can vary significantly from one country to another. As mentioned in section \ref{approach-metrics}, it is not difficult to consider mechanisms where a nation state might have an incentive to regulate the information embedded in web search results viewed by their populations. Even when governments might avoid directly manipulating information on the Internet for fear of international scrutiny, they can often indirectly down-weight certain types of search results, such as those related to sensitive political topics. Such state media control has been in the case of countries with strong state regimes like in China and Russia \citep{walker2014breaking, yang2014internet}. For such regulated topics, one can expect the information viewed regularly in the top ranked results to be less representative of the overall corpus of results, hence generating relatively lower information completeness aggregates at the country level. Figure \ref{fig4} highlights how information completeness varies with a given country's stance on freedom of media and the press. It is interesting, but at the same time intuitive, to see that information completeness follows a broad downward trend, and varies inversely with country-level press restriction scores \citep{reporters}.

Table \ref{tab31} considers this aspect through the lens of a linear model, and further reinforces that this inverse relation with press restrictions holds even if I control for a nation's search volume, its gross domestic product, its population, and the day search was made. We see that one unit increase in country-level press restriction scores (0-100) reduces aggregate information completeness by 0.28 percentage points ($p < 0.001$). Adding region fixed effects reduces the effect size magnitude to 0.17 percentage points ($p < 0.001$). This is unsurprising as press restrictions have a tendency to be spatially correlated on account of some regions being more fragile than others, but there is still a significant effect within regions at the country level. I must note that these associations are not causal, but still suggest a potential way in which our information completeness metric might be reflecting media restrictions across nations.

\bigskip

\begin{table}[H]
\begin{center}
\begin{tabular}{l c c c c c}
\toprule
 & I & II & III & IV & V \\
\midrule
Press restriction & $-0.28^{***}$ & $-0.28^{***}$ & $-0.27^{***}$ & $-0.27^{***}$ & $-0.17^{***}$ \\
                  & $(0.01)$      & $(0.01)$      & $(0.01)$      & $(0.01)$      & $(0.01)$      \\
Search volume     &               & $0.00^{***}$  & $0.00^{***}$  & $0.00^{***}$  & $0.00$        \\
                  &               & $(0.00)$      & $(0.00)$      & $(0.00)$      & $(0.00)$      \\
GDP per capita    &               &               & $0.01$        & $0.01$        & $0.10^{***}$  \\
                  &               &               & $(0.01)$      & $(0.01)$      & $(0.01)$      \\
Population        &               &               & $-0.00$       & $-0.00$       & $-0.19^{***}$ \\
                  &               &               & $(0.01)$      & $(0.01)$      & $(0.01)$      \\
\midrule
Date of Search FE & No & No & No & Yes & Yes \\
Region FE & No & No & No & No & Yes \\
\midrule
$N$ & 294098 & 294098 & 294098 & 294098 & 294098 \\
\bottomrule
\multicolumn{6}{l}{\scriptsize{$^{***}p<0.001$; $^{**}p<0.01$; $^{*}p<0.05$. All continuous variables are standardized. Effect sizes in SD units.}}
\end{tabular}
\caption{Table showing reduction in information completeness with rising press restrictions. Each observation is a date - country - search query. The dependent variable is information completeness on a 0-100 scale. Search volume, GDP per capita, and population variables are standardized. Region fixed effects include East Asia \& Pacific, Europe \& Central Asia, Latin America \& Caribbean, Middle East \& North Africa, North America, and South Asia.}
\label{tab31}
\end{center}
\end{table}

\bigskip

\begin{figure}[H]
\centering
\includegraphics[width=\textwidth,height=0.5\textheight]{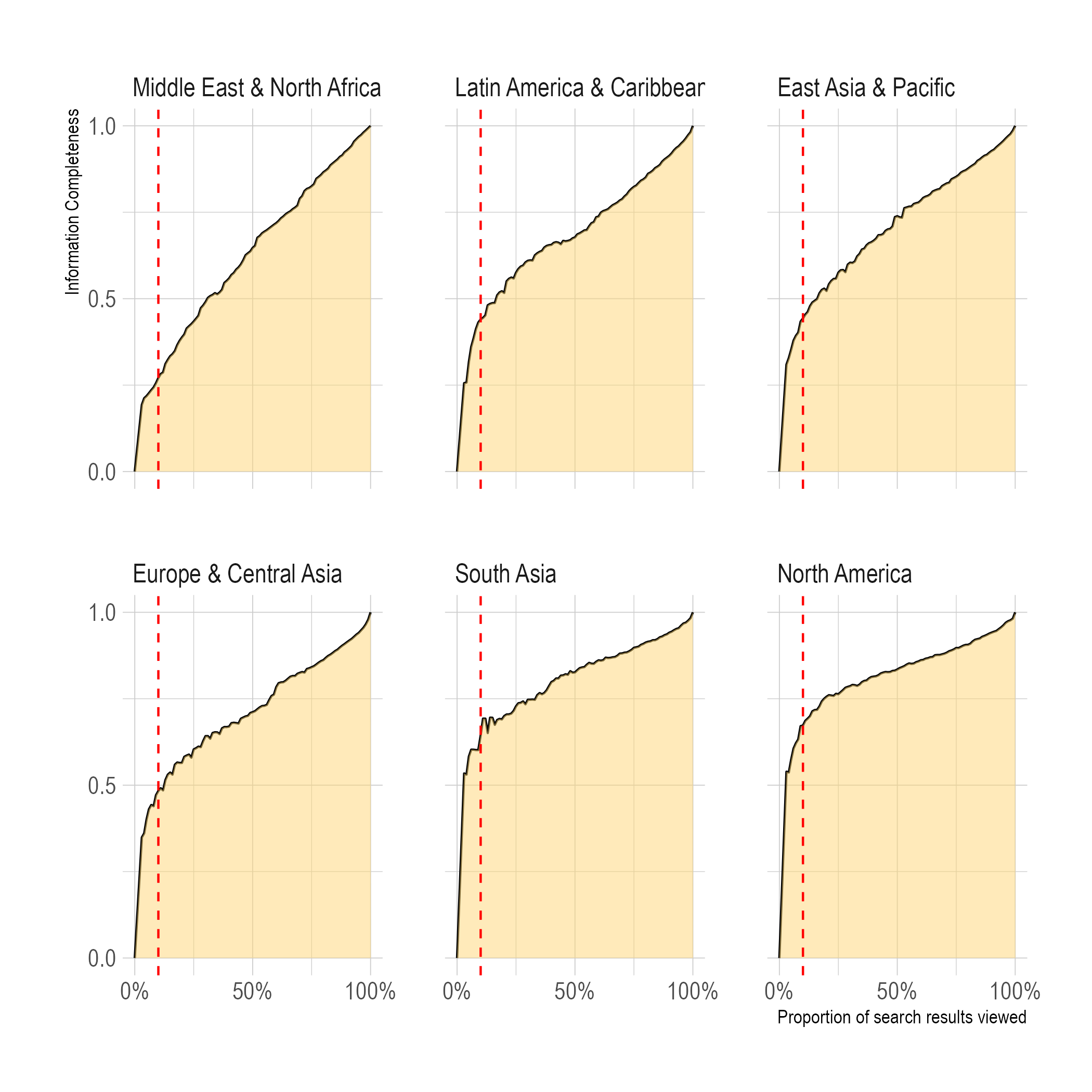}
\caption{Information Completeness curves split by geographic region and sorted by rising area under the curve.}
\label{fig5}
\end{figure}

In order to understand these regional differences further, I generate information completeness curves aggregated by six geographic regions -- Middle East \& North Africa, Latin America \& Caribbean, East Asia \& Pacific, Europe \& Central Asia, South Asia, and North America --  as shown in Figure \ref{fig5}, with the dashed line showing where the first page of search results ends. Specifically, top 100 results per query are considered, and the first page is defined as the top 10 Google search results. We see that the Middle East and North Africa region has the least complete information (the population reaches around 25\% information completeness on the first page), while North America has the most representative (the population reaches around 62\% information completeness on the first page). This is important because then people in MENA have to traverse more search results to reach a higher information completeness. Since viewing lower-ranked search results is fairly uncommon \citep{goldman2005search, introna2000shaping}, the population navigating the Internet in MENA could be settling for lower information completeness than that in North America.

\subsection{Implications of Incomplete Information Awareness}
\label{result-exp}

\begin{figure}[H]
\centering
\includegraphics[scale = 0.8]{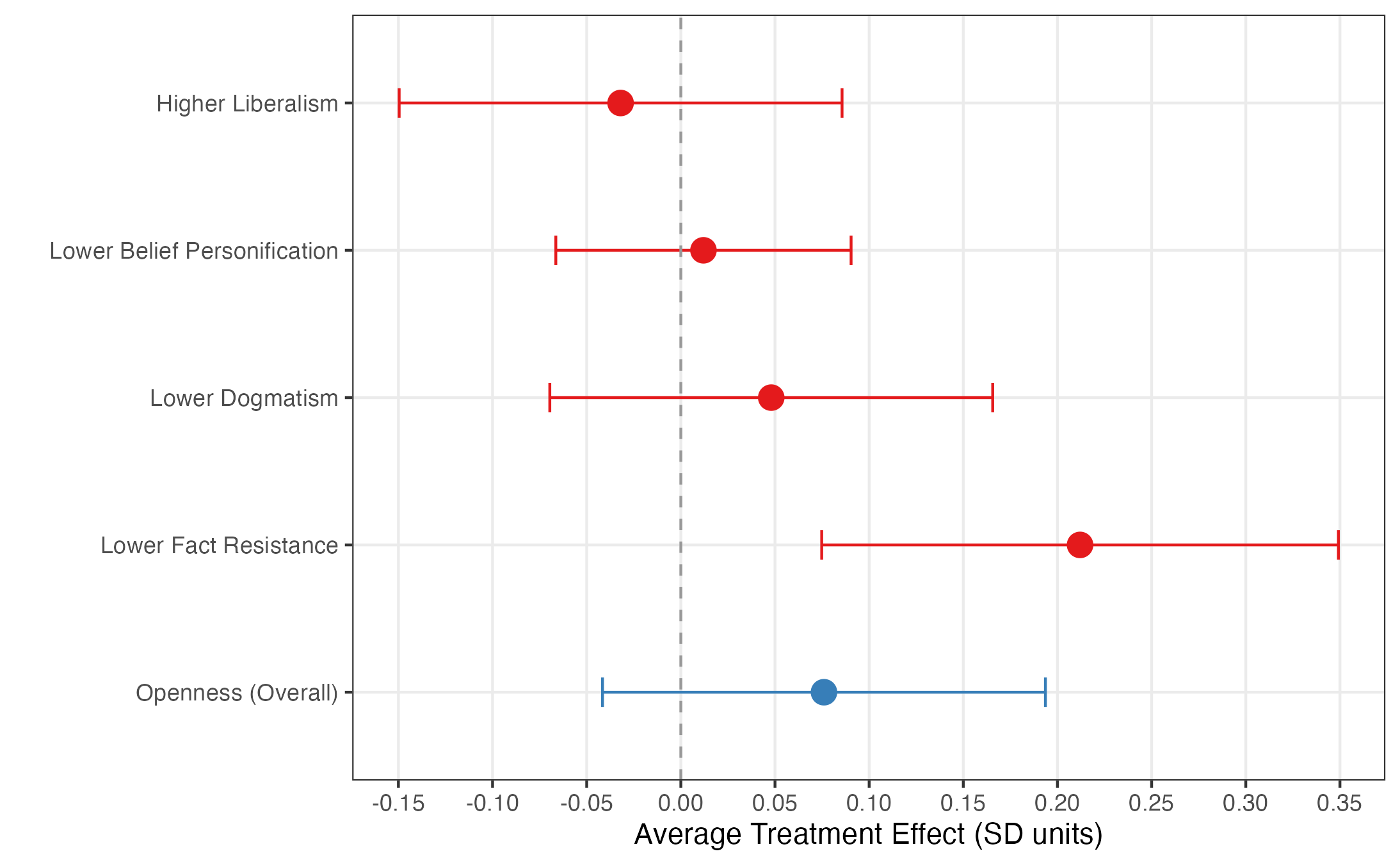}
\caption{Average treatment effects of being aware of information completeness on open-mindedness. The error bars denote 95\% confidence intervals.}
\label{fig7}
\end{figure}

Let us now shift our focus towards examining the impact of being aware of information completeness on an individual's receptiveness to novel or conflicting information, as depicted in Figure \ref{fig7}. In comparison to the control group, our findings indicate that there are slight indications suggesting that awareness of information completeness has a positive influence on overall open-mindedness, albeit not reaching statistical significance. Specifically, there is a modest increase of 0.076 standard deviation (SD) units in open-mindedness for individuals who possess this awareness ($p = 0.207$).

When we delve deeper into the subscales that measure different dimensions of open-mindedness, we observe that the positive effect is predominantly driven by the dimensions associated with knowledge-related aspects (that is, fact resistance and dogmatism). Particularly noteworthy is a statistically significant reduction in resistance to factual knowledge, which shows a substantial shift of 0.212 SD units ($p = 0.003$) due to the intervention. Additionally, we note positive but non-significant effects of the intervention on lowering individuals' tendencies to hold dogmatic beliefs, with a modest shift of 0.048 SD units ($p = 0.432$).
On assessing person-related dimensions (that is, belief personification and liberalism), we find that our treatment has only minimal and insignificant effects. There is a negligible decrease of 0.012 SD units in belief personification ($p = 0.777$), indicating that the intervention did not substantially impact the degree to which individuals attribute beliefs to themselves. Similarly, there is a slight, but statistically insignificant, decrease of 0.032 SD units in liberal thinking ($p = 1.302$), suggesting that the treatment did not have a significant influence on fostering liberal viewpoints.

Overall, our investigation reveals nuanced effects of awareness about information completeness on open-mindedness, primarily driven by improvements in knowledge-related dimensions, particularly a significant reduction in resistance to factual knowledge. However, the intervention yielded limited impact on person-related dimensions, such as belief personification and liberal thinking, and showed no significant effects on reducing dogmatic tendencies either.
\bigskip
\bigskip

\begin{figure}[H]
\centering
\includegraphics[height=0.4\textheight]{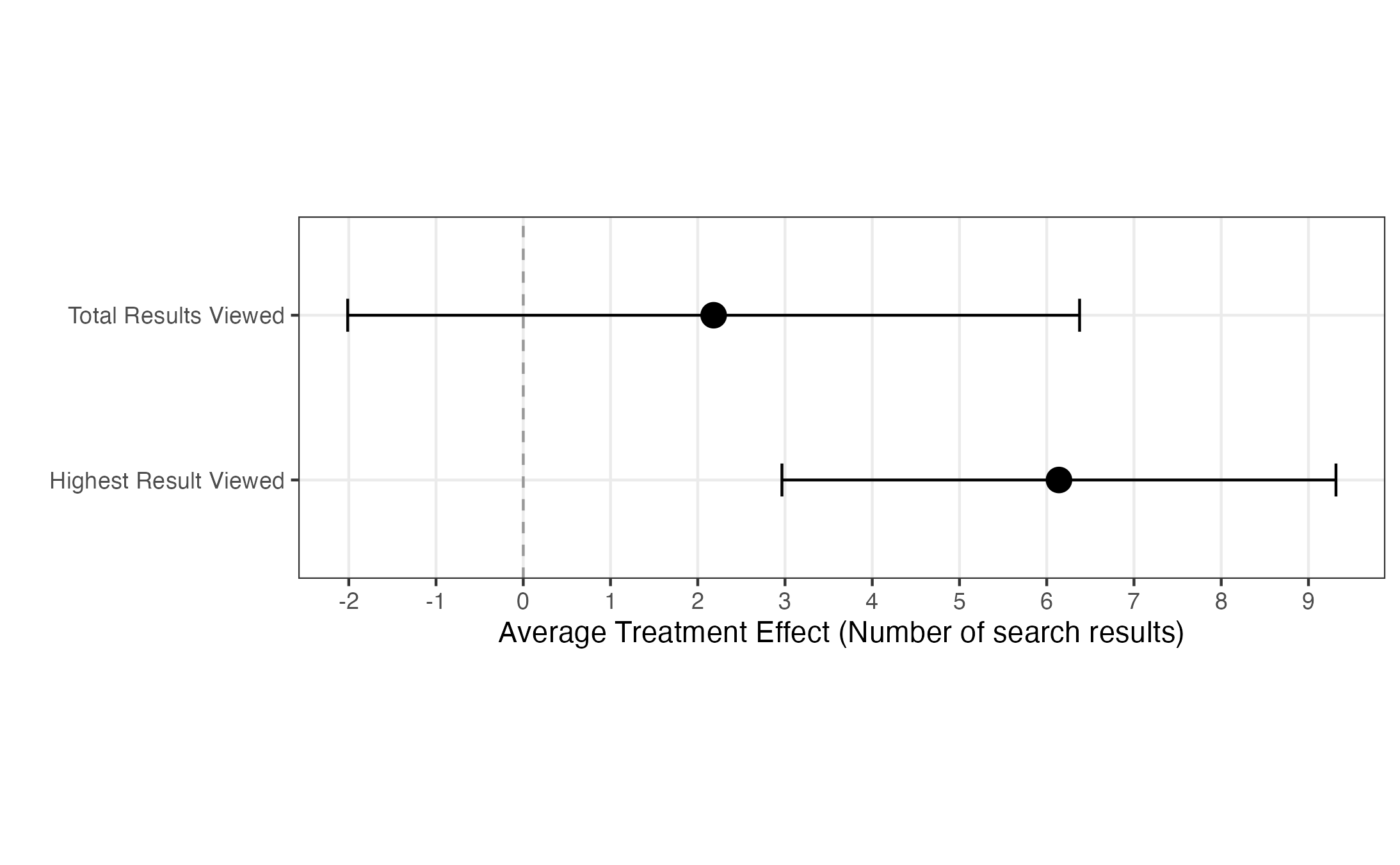}
\caption{Average treatment effects of being aware of information completeness on number of and ranks of results clicked. The error bars denote 95\% confidence intervals.}
\label{fig8}
\end{figure}

Moving ahead to our second set of outcomes, we discover that having knowledge about information completeness scores has a substantial and noteworthy impact on the extent to which individuals explore search results. This observation is supported by Figure \ref{fig8}. In comparison to the control group, participants who received the treatment exhibit a considerable shift in their browsing behavior, with the lowest-ranked result they examine being positioned 6.14 search ranks further down the list ($p < 0.001$). Notably, despite this significant alteration in their search pattern, there is little discernible distinction in the overall number of search results clicked between the treatment group and the control group participants, with the treatment group clicking on 2.182 more results on average ($p = 0.312$). Further, I find that participants in the treatment group click on results with a higher aggregate information completeness scores (higher by 7.6 percentage points, $p = 0.001$) as compared to those in the control group. 

\begin{figure}[H]
\centering
\includegraphics[height=0.4\textheight]{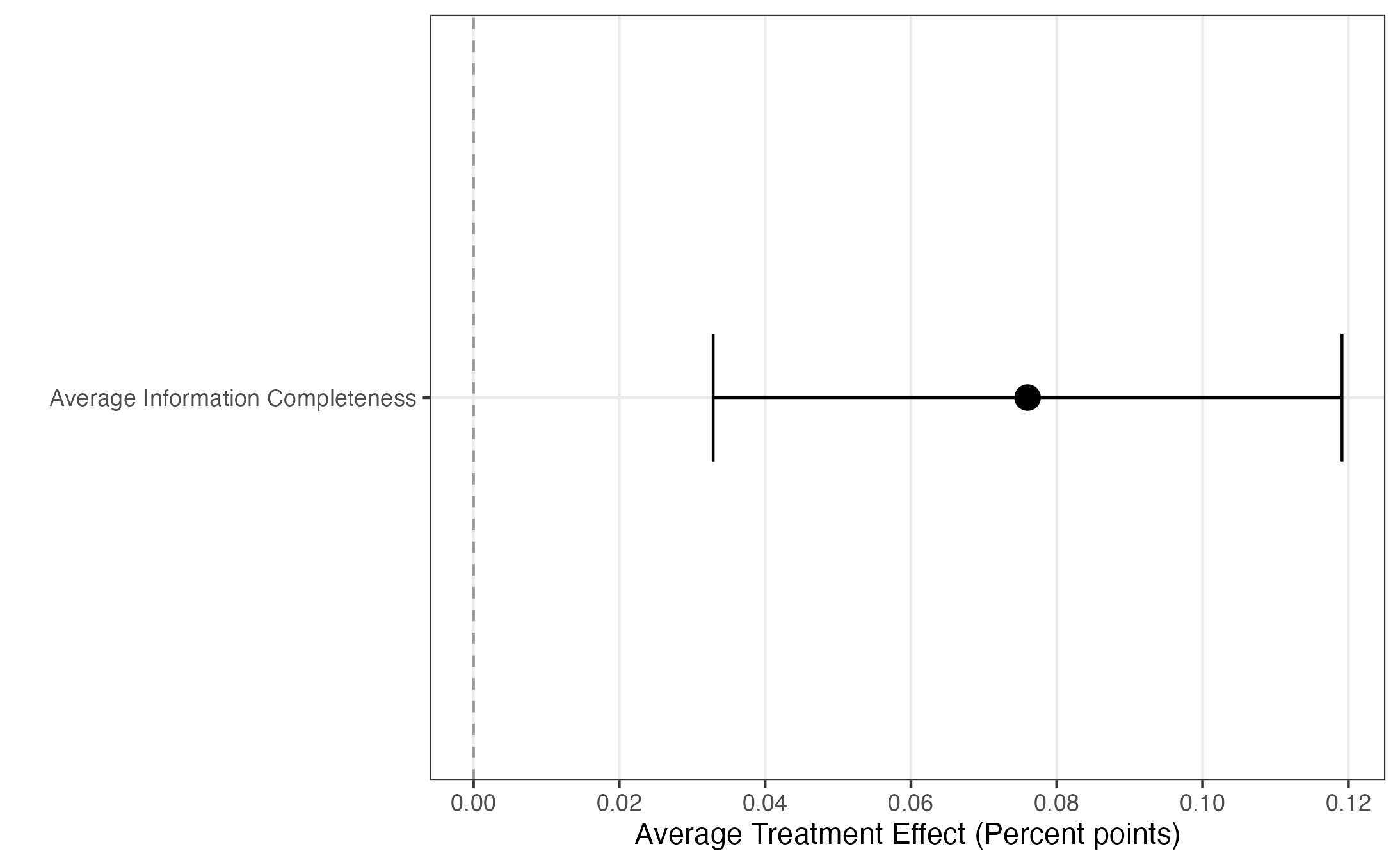}
\caption{Average treatment effects of being aware of information completeness on aggregate completeness of results clicked. The error bars denote 95\% confidence intervals.}
\label{fig9}
\end{figure}

\section{Discussion}
\label{discussion}

We are living in times of amplifying information overloads, where the sheer volume of information available to us can be overwhelming and difficult to navigate. With the proliferation of the Internet and social media, we are bombarded with an endless stream of news, opinions, and data from a variety of sources, and it can be challenging to discern what is accurate and relevant. This overload of `unknown unknowns' can lead to confusion and indecision, and it can be difficult to prioritize what is important. Additionally, the constant influx of information can be overwhelming and stressful, and it can be difficult to find time to process and reflect on it all. In these times of information overload, it is important to find ways to manage the influx of information and turn unknown unknowns to `known unknowns' at the very least.

This study fills a glaring loophole in current research by building on information retrieval and text embedding approaches to propose a novel metric that measures `information completeness' dynamically when one browses the Internet. In addition to being intuitive in terms of comparing low-dimensional vector representations of text, the metric is also validated by assessing its variation with 6.5 trillion web and news search results across 48 countries. Next, we find causal evidence that awareness of information completeness while browsing the Internet reduces resistance to factual information, consequently contributing to an increase in active open-minded thinking.
We also find that the intervention marginally might reduce a tendency for dogmatism. The fact that the treatment effect is driven by these knowledge-related dimensions from the AOT scale (and not person-related dimensions) highlights the need of person-level interventions to tackle conservatism and intolerance.
Further, in an era of personalized results where one sees more of what they already consume, we find that awareness of information completeness makes one traverse further down the information iceberg and explore lower ranked results. This in turn could be an important mechanism leading to novel and divergent knowledge sources, potentially reducing dogmatic tendencies.

I see three limitations in this research at present. The first limitation of this research arises from an aspect which is also its strength -- leveraging sentence-level text embeddings to quantify information completeness. The quality of the metric will vary directly with the quality of the text embeddings. Text embeddings are often created by training machine learning models on large datasets of text, and the quality of the embeddings can depend on the quality and diversity of the data used to train the model. If the training data is biased or does not accurately reflect the real-world distribution of the text, the resulting embeddings could be misleading. Another issue could be that text embeddings are typically created based on the statistical relationships between words and the contexts in which they appear. This means that they may not always capture the full meaning or nuances of a particular word or phrase, and may be less effective when dealing with more complex or abstract languages propagating on the Internet. That said, given the rapid improvements in transformer-based attention mechanisms to capture high-quality contextual information across hundreds of languages, I expect this limitation to be mitigated in the coming years.

A second limitation is that Actively Open-Minded Thinking or the AOT is a construct that originated in the Western world, primarily within the context of Western psychology and philosophy. Consequently, its application and interpretation might not seamlessly translate across different cultural, social, and intellectual contexts. Cultural variations in thinking styles, decision-making processes, and the value placed on open-mindedness can significantly influence how AOT is perceived and measured. For instance, in cultures where consensus and harmony are highly valued, open-mindedness might manifest differently than in cultures that value individualism and debate. Similarly, some cultures may value tradition and stability over the questioning and changing of beliefs. Therefore, while AOT provides valuable insights within its original context, its universal applicability may be limited. Researchers applying the AOT scale cross-culturally should therefore exercise caution, ensuring they take into account cultural nuances and consider potential biases in the interpretation of results.

A third issue is that the surveyed population has a limited assessing of the information completion metric. This is because a vast majority of us, myself included, are habitually used to uncritically consuming `top' search results on a daily bus. If the challenge (of information overloads) that we face has not been acknowledged, it is certainly going to be harder to propagate a successful solution.
Assessing the treatments effects of such an intervention poses substantial challenges, primarily because comprehension is a key factor in both adhering to treatment protocols and accurately reporting outcomes. If participants do not understand the treatment, they may not use it correctly, leading to unreliable results. Additionally, their ability to accurately report their experiences, symptoms, or side-effects may be compromised, further confounding the data. Misunderstanding can also lead to heightened anxiety or placebo effects, which could inadvertently influence the treatment outcomes. Therefore, it is crucial to do well on a preceding objective of educating individuals about our existence in a world of incomplete information. Once we achieve substantial success on this educational objective, it will ensure participants understand the treatment they are receiving well, including its potential effects, hence safeguarding the integrity of the final results.

In times of growing information overloads, it is important that we consistently strive to stay aware of how much stays hidden from us. Metrics like information completeness are a first step in the domain of knowing unknowns, and I see two potential directions for future research. First, an aspect not addressed by this study was assessing the causal variation in open-mindedness with the change in the `magnitude' of information completeness scores themselves. This might show us whether treatment effects vary by how complete the viewed information actually is. Second, this study is embedded in the context of Internet searches with their explicit result rankings, but can potentially be replicated in a social media context with minimal adjustments. A distinct feature of social media is its anthropomorphic nature, where every information snippet is visibly associated with a unique and active human source. This might reveal interesting implications not only for assessing information completeness, but also for how this awareness affects our behavioral choices.

In conclusion, the modern context of information overload necessitates a heightened vigilance towards the completeness and accuracy of our information. Amidst the deluge of readily available data, it is all too easy to make decisions or form opinions based on incomplete or inaccurate information. This not only impairs our individual judgement but also risks perpetuating misinformation and bias in our societies. To combat this, it's essential to adopt a critical and discerning approach to information consumption. We must continually assess our understanding, recognize gaps in our knowledge, and seek diverse and credible sources to fill those gaps. By doing so, we can transform the challenge of information overload into an opportunity for informed decision-making and enlightened discourse, ultimately fostering a more informed and discerning society.

\section*{Acknowledgements}

I thank colleagues at the Stanford Data Science Institute and the Stanford Institute for Human-centered Artificial Intelligence for funding and feedback on this study. I also thank Prashant Loyalka, Benjamin Domingue, Nicholas Haber, Sanne Smith, John Willinsky, Susan Athey, Erik Brynjolfsson, Alex Pentland, David Laitin, Jeremy Weinstein, Alex Siegel, John Chambers, Klint Kanopka, Radhika Kapoor, Ishita Ahmed, and Hansol Lee for helpful comments and suggestions.


\singlespacing
\bibliographystyle{apalike}
\bibliography{references}

\begin{thebibliography}{}

\bibitem[Abomhara and K{\o}ien, 2015]{abomhara2015cyber}
Abomhara, M. and K{\o}ien, G.~M. (2015).
\newblock Cyber security and the internet of things: vulnerabilities, threats, intruders and attacks.
\newblock {\em Journal of Cyber Security and Mobility}, pages 65--88.

\bibitem[Ali, 2020]{ali2020online}
Ali, W. (2020).
\newblock Online and remote learning in higher education institutes: A necessity in light of covid-19 pandemic.
\newblock {\em Higher education studies}, 10(3):16--25.

\bibitem[Ananny and Crawford, 2018]{ananny2018seeing}
Ananny, M. and Crawford, K. (2018).
\newblock Seeing without knowing: Limitations of the transparency ideal and its application to algorithmic accountability.
\newblock {\em new media \& society}, 20(3):973--989.

\bibitem[Anklesaria et~al., 1993]{anklesaria1993internet}
Anklesaria, F., McCahill, M., Lindner, P., Johnson, D., Torrey, D., and Albert, B. (1993).
\newblock The internet gopher protocol (a distributed document search and retrieval protocol).
\newblock Technical report.

\bibitem[Baron, 2000]{baron2000thinking}
Baron, J. (2000).
\newblock {\em Thinking and deciding}.
\newblock Cambridge University Press.

\bibitem[Becker et~al., 2007]{becker2007evaluation}
Becker, L.~B., Vlad, T., and Nusser, N. (2007).
\newblock An evaluation of press freedom indicators.
\newblock {\em International Communication Gazette}, 69(1):5--28.

\bibitem[Berners-Lee et~al., 2001]{berners2001semantic}
Berners-Lee, T., Hendler, J., and Lassila, O. (2001).
\newblock The semantic web.
\newblock {\em Scientific american}, 284(5):34--43.

\bibitem[Bok et~al., 2022]{bok2022personalized}
Bok, K., Song, J., Lim, J., and Yoo, J. (2022).
\newblock Personalized search using user preferences on social media.
\newblock {\em Electronics}, 11(19):3049.

\bibitem[Bowrey, 2005]{bowrey2005law}
Bowrey, K. (2005).
\newblock {\em Law and internet cultures}.
\newblock Cambridge University Press.

\bibitem[Brake, 2020]{brake2020lessons}
Brake, D. (2020).
\newblock Lessons from the pandemic: Broadband policy after covid-19.
\newblock Technical report, Information Technology and Innovation Foundation.

\bibitem[Bruns, 2019]{bruns2019filter}
Bruns, A. (2019).
\newblock {\em Are filter bubbles real?}
\newblock John Wiley \& Sons.

\bibitem[Brynjolfsson et~al., 2020]{brynjolfsson2020covid}
Brynjolfsson, E., Horton, J.~J., Ozimek, A., Rock, D., Sharma, G., and TuYe, H.-Y. (2020).
\newblock Covid-19 and remote work: An early look at us data.
\newblock Technical report, National Bureau of Economic Research.

\bibitem[Carri{\`e}re-Swallow and Labb{\'e}, 2013]{carriere2013nowcasting}
Carri{\`e}re-Swallow, Y. and Labb{\'e}, F. (2013).
\newblock Nowcasting with google trends in an emerging market.
\newblock {\em Journal of Forecasting}, 32(4):289--298.

\bibitem[Cavazos et~al., 2020]{cavazos2020accuracy}
Cavazos, J.~G., Phillips, P.~J., Castillo, C.~D., and O’Toole, A.~J. (2020).
\newblock Accuracy comparison across face recognition algorithms: Where are we on measuring race bias?
\newblock {\em IEEE transactions on biometrics, behavior, and identity science}, 3(1):101--111.

\bibitem[Choi and Varian, 2012]{choi2012predicting}
Choi, H. and Varian, H. (2012).
\newblock Predicting the present with google trends.
\newblock {\em Economic record}, 88:2--9.

\bibitem[Chowdhary, 2020]{chowdhary2020natural}
Chowdhary, K. (2020).
\newblock Natural language processing for word sense disambiguation and information extraction.
\newblock {\em arXiv preprint arXiv:2004.02256}.

\bibitem[Cooper et~al., 1997]{cooper1997plato}
Cooper, J.~M., Hutchinson, D.~S., et~al. (1997).
\newblock {\em Plato: complete works}.
\newblock Hackett Publishing.

\bibitem[Dahlberg, 2001]{dahlberg2001internet}
Dahlberg, L. (2001).
\newblock The internet and democratic discourse: Exploring the prospects of online deliberative forums extending the public sphere.
\newblock {\em Information, communication \& society}, 4(4):615--633.

\bibitem[Daniel, 2020]{daniel2020education}
Daniel, S.~J. (2020).
\newblock Education and the covid-19 pandemic.
\newblock {\em Prospects}, 49(1):91--96.

\bibitem[De~Gregorio and Stremlau, 2020]{de2020internet}
De~Gregorio, G. and Stremlau, N. (2020).
\newblock Internet shutdowns and the limits of law.

\bibitem[Devlin et~al., 2018]{DBLP:journals/corr/abs-1810-04805}
Devlin, J., Chang, M., Lee, K., and Toutanova, K. (2018).
\newblock {BERT:} pre-training of deep bidirectional transformers for language understanding.
\newblock {\em CoRR}, abs/1810.04805.

\bibitem[Einstein et~al., 1931]{einstein1931knowledge}
Einstein, A., Tolman, R.~C., and Podolsky, B. (1931).
\newblock Knowledge of past and future in quantum mechanics.
\newblock {\em Physical Review}, 37(6):780.

\bibitem[Everett, 2013]{everett201312}
Everett, J.~A. (2013).
\newblock The 12 item social and economic conservatism scale (secs).
\newblock {\em PloS one}, 8(12):e82131.

\bibitem[Fettweis, 2014]{fettweis2014tactile}
Fettweis, G.~P. (2014).
\newblock The tactile internet: Applications and challenges.
\newblock {\em IEEE Vehicular Technology Magazine}, 9(1):64--70.

\bibitem[Gerbaudo, 2017]{gerbaudo2017cyber}
Gerbaudo, P. (2017).
\newblock From cyber-autonomism to cyber-populism: An ideological history of digital activism.
\newblock {\em TripleC: Communication, Capitalism \& Critique}, 15(2):477--489.

\bibitem[Goldberg and Levy, 2014]{goldberg2014word2vec}
Goldberg, Y. and Levy, O. (2014).
\newblock word2vec explained: deriving mikolov et al.'s negative-sampling word-embedding method.
\newblock {\em arXiv preprint arXiv:1402.3722}.

\bibitem[Goldman, 2005]{goldman2005search}
Goldman, E. (2005).
\newblock Search engine bias and the demise of search engine utopianism.
\newblock {\em Yale JL \& Tech.}, 8:188.

\bibitem[Groeling, 2013]{groeling2013media}
Groeling, T. (2013).
\newblock Media bias by the numbers: Challenges and opportunities in the empirical study of partisan news.
\newblock {\em Annual Review of Political Science}, 16(1):129--151.

\bibitem[Guy and Carmel, 2011]{guy2011social}
Guy, I. and Carmel, D. (2011).
\newblock Social recommender systems.
\newblock In {\em Proceedings of the 20th international conference companion on World wide web}, pages 283--284.

\bibitem[Hague and Loader, 1999]{hague1999digital}
Hague, B.~N. and Loader, B. (1999).
\newblock {\em Digital democracy: Discourse and decision making in the information age}.
\newblock Psychology Press.

\bibitem[Haran et~al., 2013]{haran2013role}
Haran, U., Ritov, I., and Mellers, B.~A. (2013).
\newblock The role of actively open-minded thinking in information acquisition, accuracy, and calibration.
\newblock {\em Judgment and Decision making}, 8(3):188--201.

\bibitem[Harris, 2022]{harris2022happiness}
Harris, R. (2022).
\newblock {\em The happiness trap: How to stop struggling and start living}.
\newblock Shambhala Publications.

\bibitem[Hassani et~al., 2021]{hassani2021escaping}
Hassani, A., Walton, S., Shah, N., Abuduweili, A., Li, J., and Shi, H. (2021).
\newblock Escaping the big data paradigm with compact transformers.
\newblock {\em arXiv preprint arXiv:2104.05704}.

\bibitem[Herlocker et~al., 2004]{herlocker2004evaluating}
Herlocker, J.~L., Konstan, J.~A., Terveen, L.~G., and Riedl, J.~T. (2004).
\newblock Evaluating collaborative filtering recommender systems.
\newblock {\em ACM Transactions on Information Systems (TOIS)}, 22(1):5--53.

\bibitem[Hermes, 2006]{hermes2006citizenship}
Hermes, J. (2006).
\newblock Citizenship in the age of the internet.
\newblock {\em European Journal of Communication}, 21(3):295--309.

\bibitem[Hofman et~al., 2021]{hofman2021integrating}
Hofman, J.~M., Watts, D.~J., Athey, S., Garip, F., Griffiths, T.~L., Kleinberg, J., Margetts, H., Mullainathan, S., Salganik, M.~J., Vazire, S., et~al. (2021).
\newblock Integrating explanation and prediction in computational social science.
\newblock {\em Nature}, 595(7866):181--188.

\bibitem[ILS, 2022]{stats}
ILS (2022).
\newblock Internet live stats - internet usage \& social media statistics.

\bibitem[Introna and Nissenbaum, 2000]{introna2000shaping}
Introna, L.~D. and Nissenbaum, H. (2000).
\newblock Shaping the web: Why the politics of search engines matters.
\newblock {\em The information society}, 16(3):169--185.

\bibitem[Irie et~al., 2019]{irie2019language}
Irie, K., Zeyer, A., Schl{\"u}ter, R., and Ney, H. (2019).
\newblock Language modeling with deep transformers.
\newblock {\em arXiv preprint arXiv:1905.04226}.

\bibitem[Joulin et~al., 2016]{joulin2016fasttext}
Joulin, A., Grave, E., Bojanowski, P., Douze, M., J{\'e}gou, H., and Mikolov, T. (2016).
\newblock Fasttext. zip: Compressing text classification models.
\newblock {\em arXiv preprint arXiv:1612.03651}.

\bibitem[Kulshrestha et~al., 2017]{kulshrestha2017quantifying}
Kulshrestha, J., Eslami, M., Messias, J., Zafar, M.~B., Ghosh, S., Gummadi, K.~P., and Karahalios, K. (2017).
\newblock Quantifying search bias: Investigating sources of bias for political searches in social media.
\newblock In {\em Proceedings of the 2017 ACM Conference on Computer Supported Cooperative Work and Social Computing}, pages 417--432.

\bibitem[Kusner et~al., 2015]{kusner2015word}
Kusner, M., Sun, Y., Kolkin, N., and Weinberger, K. (2015).
\newblock From word embeddings to document distances.
\newblock In {\em International conference on machine learning}, pages 957--966. PMLR.

\bibitem[Lazer et~al., 2009]{lazer2009computational}
Lazer, D., Pentland, A., Adamic, L., Aral, S., Barab{\'a}si, A.-L., Brewer, D., Christakis, N., Contractor, N., Fowler, J., Gutmann, M., et~al. (2009).
\newblock Computational social science.
\newblock {\em Science}, 323(5915):721--723.

\bibitem[Lazer et~al., 2020]{lazer2020computational}
Lazer, D.~M., Pentland, A., Watts, D.~J., Aral, S., Athey, S., Contractor, N., Freelon, D., Gonzalez-Bailon, S., King, G., Margetts, H., et~al. (2020).
\newblock Computational social science: Obstacles and opportunities.
\newblock {\em Science}, 369(6507):1060--1062.

\bibitem[Ledford, 2019]{ledford2019millions}
Ledford, H. (2019).
\newblock Millions affected by racial bias in health-care algorithm.
\newblock {\em Nature}, 574(31):2.

\bibitem[Lee, 2011]{lee2011google}
Lee, M. (2011).
\newblock Google ads and the blindspot debate.
\newblock {\em Media, Culture \& Society}, 33(3):433--447.

\bibitem[Leeson, 2008]{leeson2008media}
Leeson, P.~T. (2008).
\newblock Media freedom, political knowledge, and participation.
\newblock {\em Journal of Economic Perspectives}, 22(2):155--169.

\bibitem[Levy and Goldberg, 2014]{levy2014dependency}
Levy, O. and Goldberg, Y. (2014).
\newblock Dependency-based word embeddings.
\newblock In {\em Proceedings of the 52nd Annual Meeting of the Association for Computational Linguistics (Volume 2: Short Papers)}, pages 302--308.

\bibitem[Liu et~al., 2019]{DBLP:journals/corr/abs-1907-11692}
Liu, Y., Ott, M., Goyal, N., Du, J., Joshi, M., Chen, D., Levy, O., Lewis, M., Zettlemoyer, L., and Stoyanov, V. (2019).
\newblock Roberta: {A} robustly optimized {BERT} pretraining approach.
\newblock {\em CoRR}, abs/1907.11692.

\bibitem[Livingston, 2007]{livingston2007brewster}
Livingston, J. (2007).
\newblock Brewster kahle: Founder, wais, internet archive, alexa internet.
\newblock {\em Founders at Work: Stories of Startups’ Early Days}, pages 265--280.

\bibitem[Lu and Da~Xu, 2018]{lu2018internet}
Lu, Y. and Da~Xu, L. (2018).
\newblock Internet of things (iot) cybersecurity research: A review of current research topics.
\newblock {\em IEEE Internet of Things Journal}, 6(2):2103--2115.

\bibitem[Mardikian, 1994]{mardikian1994use}
Mardikian, J. (1994).
\newblock How to use veronica to find information on the internet.
\newblock {\em The Reference Librarian}, 19(41-42):37--45.

\bibitem[Mellon, 2014]{mellon2014internet}
Mellon, J. (2014).
\newblock Internet search data and issue salience: The properties of google trends as a measure of issue salience.
\newblock {\em Journal of Elections, Public Opinion \& Parties}, 24(1):45--72.

\bibitem[Nuti et~al., 2014]{nuti2014use}
Nuti, S.~V., Wayda, B., Ranasinghe, I., Wang, S., Dreyer, R.~P., Chen, S.~I., and Murugiah, K. (2014).
\newblock The use of google trends in health care research: a systematic review.
\newblock {\em PloS one}, 9(10):e109583.

\bibitem[Obermeyer et~al., 2019]{obermeyer2019dissecting}
Obermeyer, Z., Powers, B., Vogeli, C., and Mullainathan, S. (2019).
\newblock Dissecting racial bias in an algorithm used to manage the health of populations.
\newblock {\em Science}, 366(6464):447--453.

\bibitem[Paasonen, 2021]{paasonen2021dependent}
Paasonen, S. (2021).
\newblock {\em Dependent, distracted, bored: Affective formations in networked media}.
\newblock MIT Press.

\bibitem[Page et~al., 1998]{page1998pagerank}
Page, L., Brin, S., Motwani, R., and Winograd, T. (1998).
\newblock The pagerank citation ranking: Bring order to the web.
\newblock Technical report, technical report, Stanford University.

\bibitem[Palan and Schitter, 2018]{palan2018prolific}
Palan, S. and Schitter, C. (2018).
\newblock Prolific. ac—a subject pool for online experiments.
\newblock {\em Journal of Behavioral and Experimental Finance}, 17:22--27.

\bibitem[Palangi et~al., 2016]{palangi2016deep}
Palangi, H., Deng, L., Shen, Y., Gao, J., He, X., Chen, J., Song, X., and Ward, R. (2016).
\newblock Deep sentence embedding using long short-term memory networks: Analysis and application to information retrieval.
\newblock {\em IEEE/ACM Transactions on Audio, Speech, and Language Processing}, 24(4):694--707.

\bibitem[Pennington et~al., 2014]{pennington2014glove}
Pennington, J., Socher, R., and Manning, C.~D. (2014).
\newblock Glove: Global vectors for word representation.
\newblock In {\em Proceedings of the 2014 conference on empirical methods in natural language processing (EMNLP)}, pages 1532--1543.

\bibitem[Reimers and Gurevych, 2019]{DBLP:journals/corr/abs-1908-10084}
Reimers, N. and Gurevych, I. (2019).
\newblock Sentence-bert: Sentence embeddings using siamese bert-networks.
\newblock {\em CoRR}, abs/1908.10084.

\bibitem[Roozenbeek et~al., 2020]{roozenbeek2020susceptibility}
Roozenbeek, J., Schneider, C.~R., Dryhurst, S., Kerr, J., Freeman, A.~L., Recchia, G., Van Der~Bles, A.~M., and Van Der~Linden, S. (2020).
\newblock Susceptibility to misinformation about covid-19 around the world.
\newblock {\em Royal Society open science}, 7(10):201199.

\bibitem[RSF, 2021]{reporters}
RSF (2021).
\newblock 2021 world press freedom index.

\bibitem[Rydning et~al., 2018]{rydning2018digitization}
Rydning, D. R.-J. G.-J., Reinsel, J., and Gantz, J. (2018).
\newblock The digitization of the world from edge to core.
\newblock {\em Framingham: International Data Corporation}, 16.

\bibitem[Sangers et~al., 2013]{sangers2013semantic}
Sangers, J., Frasincar, F., Hogenboom, F., and Chepegin, V. (2013).
\newblock Semantic web service discovery using natural language processing techniques.
\newblock {\em Expert Systems with Applications}, 40(11):4660--4671.

\bibitem[Schafer et~al., 2007]{schafer2007collaborative}
Schafer, J.~B., Frankowski, D., Herlocker, J., and Sen, S. (2007).
\newblock Collaborative filtering recommender systems.
\newblock {\em The adaptive web: methods and strategies of web personalization}, pages 291--324.

\bibitem[Schwartz, 1996]{schwartz1996netactivism}
Schwartz, E. (1996).
\newblock {\em Netactivism: How citizens use the Internet}.
\newblock O'Reilly \& Associates, Inc.

\bibitem[Schwartz et~al., 1992]{schwartz1992comparison}
Schwartz, M.~F., Emtage, A., Kahle, B., and Neuman, B.~C. (1992).
\newblock A comparison of internet resource discovery approaches.
\newblock {\em Computing Systems}, 5(4):461--493.

\bibitem[Seymour et~al., 2011]{seymour2011history}
Seymour, T., Frantsvog, D., Kumar, S., et~al. (2011).
\newblock History of search engines.
\newblock {\em International Journal of Management \& Information Systems (IJMIS)}, 15(4):47--58.

\bibitem[Sherstinsky, 2020]{sherstinsky2020fundamentals}
Sherstinsky, A. (2020).
\newblock Fundamentals of recurrent neural network (rnn) and long short-term memory (lstm) network.
\newblock {\em Physica D: Nonlinear Phenomena}, 404:132306.

\bibitem[Stanovich and West, 2007]{stanovich2007natural}
Stanovich, K.~E. and West, R.~F. (2007).
\newblock Natural myside bias is independent of cognitive ability.
\newblock {\em Thinking \& Reasoning}, 13(3):225--247.

\bibitem[Susskind, 2018]{susskind2018future}
Susskind, J. (2018).
\newblock {\em Future politics: Living together in a world transformed by tech}.
\newblock Oxford University Press.

\bibitem[Svedholm-H{\"a}kkinen and Lindeman, 2018]{svedholm2018actively}
Svedholm-H{\"a}kkinen, A.~M. and Lindeman, M. (2018).
\newblock Actively open-minded thinking: development of a shortened scale and disentangling attitudes towards knowledge and people.
\newblock {\em Thinking \& Reasoning}, 24(1):21--40.

\bibitem[Swire-Thompson and Lazer, 2019]{swire2019public}
Swire-Thompson, B. and Lazer, D. (2019).
\newblock Public health and online misinformation: challenges and recommendations.
\newblock {\em Annual review of public health}, 41:433--451.

\bibitem[Taleb, 2007]{taleb2007black}
Taleb, N.~N. (2007).
\newblock Black swans and the domains of statistics.
\newblock {\em The american statistician}, 61(3):198--200.

\bibitem[Tennant et~al., 1994]{tennant1994crossing}
Tennant, R., Ober, J., and Lipow, A.~G. (1994).
\newblock {\em Crossing the Internet threshold: an instructional handbook}.
\newblock ERIC.

\bibitem[Vaswani et~al., 2017]{DBLP:journals/corr/VaswaniSPUJGKP17}
Vaswani, A., Shazeer, N., Parmar, N., Uszkoreit, J., Jones, L., Gomez, A.~N., Kaiser, L., and Polosukhin, I. (2017).
\newblock Attention is all you need.
\newblock {\em CoRR}, abs/1706.03762.

\bibitem[Vogels et~al., 2022]{vogels2022teens}
Vogels, E.~A., Gelles-Watnick, R., and Massarat, N. (2022).
\newblock Teens, social media and technology 2022.

\bibitem[Vosoughi et~al., 2018]{vosoughi2018spread}
Vosoughi, S., Roy, D., and Aral, S. (2018).
\newblock The spread of true and false news online.
\newblock {\em science}, 359(6380):1146--1151.

\bibitem[Walker and Orttung, 2014]{walker2014breaking}
Walker, C. and Orttung, R.~W. (2014).
\newblock Breaking the news: The role of state-run media.
\newblock {\em Journal of Democracy}, 25(1):71--85.

\bibitem[West and Bergstrom, 2021]{west2021misinformation}
West, J.~D. and Bergstrom, C.~T. (2021).
\newblock Misinformation in and about science.
\newblock {\em Proceedings of the National Academy of Sciences}, 118(15):e1912444117.

\bibitem[Wilks et~al., 2009]{wilks2009natural}
Wilks, Y., Brewster, C., et~al. (2009).
\newblock Natural language processing as a foundation of the semantic web.
\newblock {\em Foundations and Trends{\textregistered} in Web Science}, 1(3--4):199--327.

\bibitem[Wok and Mohamed, 2017]{wok2017internet}
Wok, S. and Mohamed, S. (2017).
\newblock Internet and social media in malaysia: Development, challenges and potentials.
\newblock In {\em The evolution of media communication}. IntechOpen.

\bibitem[Woloszko, 2020]{woloszko2020tracking}
Woloszko, N. (2020).
\newblock Tracking activity in real time with google trends.

\bibitem[Xia et~al., 2019]{xia2019efficient}
Xia, H., Hu, C.-q., Xiao, F., Cheng, X.-g., and Pan, Z.-k. (2019).
\newblock An efficient social-like semantic-aware service discovery mechanism for large-scale internet of things.
\newblock {\em Computer Networks}, 152:210--220.

\bibitem[Yang, 2014]{yang2014internet}
Yang, G. (2014).
\newblock Internet activism \& the party-state in china.
\newblock {\em Daedalus}, 143(2):110--123.

\bibitem[Yoganarasimhan, 2020]{yoganarasimhan2020search}
Yoganarasimhan, H. (2020).
\newblock Search personalization using machine learning.
\newblock {\em Management Science}, 66(3):1045--1070.

\bibitem[Yu et~al., 2019]{yu2019review}
Yu, Y., Si, X., Hu, C., and Zhang, J. (2019).
\newblock A review of recurrent neural networks: Lstm cells and network architectures.
\newblock {\em Neural computation}, 31(7):1235--1270.

\bibitem[Yue et~al., 2012]{yue2012analysis}
Yue, X., Di, G., Yu, Y., Wang, W., and Shi, H. (2012).
\newblock Analysis of the combination of natural language processing and search engine technology.
\newblock {\em Procedia Engineering}, 29:1636--1639.

\bibitem[Yue et~al., 2010]{yue2010beyond}
Yue, Y., Patel, R., and Roehrig, H. (2010).
\newblock Beyond position bias: Examining result attractiveness as a source of presentation bias in clickthrough data.
\newblock In {\em Proceedings of the 19th international conference on World wide web}, pages 1011--1018.

\bibitem[Zhou et~al., 2012]{zhou2012state}
Zhou, X., Xu, Y., Li, Y., Josang, A., and Cox, C. (2012).
\newblock The state-of-the-art in personalized recommender systems for social networking.
\newblock {\em Artificial Intelligence Review}, 37:119--132.

\bibitem[Ziakis et~al., 2019]{ziakis2019important}
Ziakis, C., Vlachopoulou, M., Kyrkoudis, T., and Karagkiozidou, M. (2019).
\newblock Important factors for improving google search rank.
\newblock {\em Future internet}, 11(2):32.

\end{thebibliography}

\end{document}